\documentclass[11pt]{article}
\usepackage[textwidth=16cm, textheight= 22cm]{geometry}

\usepackage{lineno,hyperref}
\modulolinenumbers[1]

\usepackage{lineno}
\usepackage[figuresright]{rotating}
\usepackage[numbers]{natbib}
\usepackage{graphics,multirow}
\usepackage{amsmath}
\usepackage{amssymb}
\usepackage{epsfig}
\usepackage{amsfonts}
\usepackage{color}
\usepackage{graphicx}
\usepackage[figuresright]{rotating}
\usepackage{array}
\usepackage{upgreek}

\renewcommand{\baselinestretch}{1.2}

\newcommand{\speq}{ \; = \; }

\newcommand{\spp}{ \, + \, }
\newcommand{\spm}{ \, - \, }

\def\eeqn{\end{equation}}

\def\el{{\ell}}

\def\d2el{\partial^2\el}

\def\el{{\ell}}

\def\d2el{\partial^2\el}

\def\newpage{\par\vfil\eject}
\def\beq{\begin{equation}}
\def\eeq{\end{equation}}

\def\csta{ \boldsymbol{\tau} }
\def\cshat{{\widehat \csta}}

\def\sepsilon2{\sigma_\epsilon^2}
\def\see2{\sigma_{\epsilon E}^2}
\def\sec2{\sigma_{\epsilon C}^2}
\def\lcone {\tau_1}
\def\lctwo {\tau_2 }

\def\c2sta {\lctwo}
\def\lcthree {\tau_3 }
\def\lcfour {\tau_4 }

\def\ut{\underline{t}}
\def\uu{\underline{u}}
\def\uc{\underline{T}}
\def\ubhat{\hat\ubeta}

\def\und{\underline}
\def\Yhat{\hat Y}

\def\sigz2{\sigma^2}
\def\szhat{\widehat{\sigz2}}

\def\bhat{\widehat{\beta}}

\def\utheta{\underline{\theta}}

\def\ubeta{\underline{\beta}}
\def\ubhat{\underline{\bhat}}
\def\uy{\underline{y}}

\def\ux{\underline{x}}

\def\sz2{ \sigma ^ 2 }
\def\sig2{\sigma ^2}
\def\sc2{ \sigma_c^ 2 }
\def\se2{ \sigma_e^ 2 }
\def\gc2{ \gamma_C  }
\def\gam2{\gamma }
\def\ge2{ \gamma_E }

\newcolumntype{L}[1]{>{\raggedright\let\newline\\\arraybackslash\hspace{0pt}}m{#1}}
\newcolumntype{C}[1]{>{\centering\let\newline\\\arraybackslash\hspace{0pt}}m{#1}}
\newcolumntype{R}[1]{>{\raggedleft\let\newline\\\arraybackslash\hspace{0pt}}m{#1}}

\begin{document}

\title{Iterative Method for Tuning Complex Simulation Code}

\author{ \bf{Yun Am Seo}${}^{a}$,
       \\  $~$\bf{Youngsaeng Lee}${}^{b}$, $~$
       and $~$\bf{Jeong-Soo Park}${}^{c, *}$
        \\ \\
      \small\it a: AI Weather Forecast Res Team, National Inst of Meteorological Science, Korea\\
      \small\it b: Digital Transformation Department, Korea Electric Power Corporation, Korea\\
        \small\it c: Department of Statistics, Chonnam National University, Gwangju 500-757, Korea\\
     \small\it *: Corresponding author, E-mail: jspark@jnu.ac.kr, Tel: +82-62-530-3445 
 }

\maketitle 

\begin{abstract}
Tuning a complex simulation code refers to the process of improving the agreement of a
code calculation with respect to a set of experimental data by adjusting
parameters implemented in the code. This process belongs to the class of inverse problems or model calibration.
For this problem, the approximated nonlinear least squares (ANLS) method
based on a Gaussian process (GP) metamodel has been employed by some researchers. A
potential drawback of the ANLS method is that the metamodel is built only once
and not updated thereafter. To address this difficulty, we propose an iterative algorithm in this study.
In the proposed algorithm, the parameters of the simulation code and GP metamodel
are alternatively re-estimated and updated by maximum likelihood estimation
and the ANLS method. This algorithm uses both computer and experimental
data repeatedly until convergence. A study using toy-models including inexact computer code with bias terms reveals that the
proposed algorithm performs better than the ANLS method and the conditional-likelihood-based approach. 
Finally, an application to a nuclear fusion simulation
code is illustrated.\footnote[1]{{\it Communications in Statistics -- Simulation and Computation}, 2020. doi:10.1080/03610918.2020.1728317}
\end{abstract}

{\noindent {Keywords: Best linear unbiased prediction; Calibration; Computer experiments; Inexact computer model; Kriging;
Numerical optimization.}}


\section{Introduction}

 Modern computer simulation codes contain various unknown parameters.
Assuming the validity of the simulation code, we can adjust or estimate such
parameters using the nonlinear least squares estimation (NLSE) method, which
minimizes the sum of the squared differences between computer responses and real
observations. This procedure is called calibration (e.g., Kennedy and O'Hagan 2001; Higdon et al. 2008; Tuo and Wu 2018)
 or code tuning (e.g., Cox, Park, and Singer 2001; Kumar 2015).
It is formally defined as the process of improving the agreement of a code
calculation or set of code calculations with respect to a chosen and fixed set of
experimental data via adjustment of the parameters implemented in the
code (Trucano et al. 2006). Han, Santner, and Rawlinson (2009) differentiated between tuning parameter and
calibration parameter. In this study, however, the two parameters are treated as the
same, henceforth it is referred to as the tuning parameter.

 If a simulation program is complex, with one execution requiring
several hours, the NLSE method may not be computationally feasible. In this
case, a statistical metamodel can be built to approximate the unknown
functional relationship between a set of controllable input variables and a
simulated response variable. This metamodel is then employed in place of the
original simulation code in the NLSE method, making the problem solvable and
computationally feasible. This method was described by Cox, Park, and Singer (2001), where a Gaussian process (GP) model was employed as the metamodel of a
complex simulation code. It is called the approximated NLS
(ANLS) method. As alternatives to the ANLS, Cox, Park, and Singer (2001) proposed likelihood-based methods.
 Kennedy and O'Hagan (2001) introduced a full Bayesian calibration method. See, for example, Higdon et al. (2004), Henderson et al. (2009), Guillas, Glover, and Malki-Epshtein (2014), and Pratola and Higdon (2016) for various Bayesian calibrations.

A potential drawback of the ANLS method is that the metamodel is built only
once and not updated thereafter; thus, the computer data are no longer used live. To address this,
 in the present report, an iterative algorithm is proposed, in which the likelihood function is {\it maximized}, and the squared
 distance is {\it minimized} iteratively until convergence is achieved.
  That is, the tuning parameters of the simulation code and model parameters of the GP are repeatedly re-estimated and updated.
The parameters of the GP model are estimated from updated
combined data using the maximum likelihood approach. Then, the
tuning parameters are re-estimated using the ANLS method. These two
optimizations are iteratively executed until convergence is achieved. We call this method the ``Max-min" algorithm.

The remainder of this report proceeds as follows. Section 2 presents a GP
model to approximate a complex simulation code and real experimental
data. Section 3 describes the calibration methods, including our proposed method, and the
calculation of standard errors. Section 4 presents a toy-model simulation study.
Section 5 applies the proposed methods to a nuclear fusion simulator, and 
Section 6 summarizes the results and provides suggestions for future research.
Some details of this study are provided in the Supplemental Material.

\section{Gaussian process model as a surrogate}

For the metamodel of the simulation code, we use a GP or a spatial regression model that
treats response
 $y(\ux)$ as a realization of a random function superimposed
on a regression model:
\begin{linenomath}
\begin{equation} \label{gpm}
Y(\ux) ~=~\sum_{j=1}^k \beta _j f _ j
(\ux) ~+~ Z(\ux) 
\end{equation}
\end{linenomath}
\begin{linenomath}
where $f$s are known functions, and
$\beta$s are unknown regression coefficients. Here, the random process $Z(.)$,
which represents a departure from the assumed linear model,
 is assumed to be
a GP with mean zero and covariance
\begin{equation}
 cov(\ut,\uu) ~=~ V(\ut,\uu) ~=~ \sz2 ~R(\ut,\uu)
\end{equation}
between $Z(\ut)$ and $Z(\uu)$ for $\ut=( t_1 ,...,t_d ), ~\uu=( u_1 ,...,u_d )$,
\end{linenomath}
where $\sz2$ represents the process variance (a scale factor), and $R(\ut,\uu)$ is the
correlation function. When the response of a computer code is stochastic, the random component term
$\epsilon$ is added to the model (\ref{gpm}).
However, we do not include $\epsilon$ in this study because a computer code is assumed to be deterministic.

Some possible choices for the correlation function are obtained from the Gaussian correlations denoted by
\begin{linenomath}
\beq \label{eqno2-2}
 R(\ut,\uu)~=~ exp ~[-\theta \sum _{i=1}^d ~ |~ t_i - u_i | ^2 ],
\eeq
\end{linenomath}
where $\theta \geq 0 $. These are special cases of a power exponential family with a power of 2.
 The non-negative parameter $ \theta $ determines the covariance
structure of $Z$: a small $\theta$ reflects high correlations between nearby
observations, whereas a large $\theta$ reflects low nearby correlations. One may
consider a different version of (\ref{eqno2-2}) by taking several $\theta$ values as follows:
\begin{linenomath}
\beq \label{eqno2-3}
 R(\ut,\uu)~=~ exp ~[- \sum _{i=1}^d ~\theta_i \; |~ t_i - u_i | ^2 ],
\eeq
\end{linenomath}
where $\theta_i  \geq 0$, for $i=1,2,\cdots, d$. This is a legitimate separable correlation function because it is a product of valid correlation functions.
 We refer the readers to Santner, Williams, and Notz (2018) for more information on this
GP model and its application to the design and analysis of
computer experiments.

Once the data have been collected at the design sites or at ``training" inputs $X$, the parameters are estimated via the maximum likelihood estimation (MLE) method and are then plugged in to predict $y(\underline x_0)$ as in (\ref{pred1}), where 
$\underline x_0$ is an untried site or a ``test" input. This prediction is called Kriging. The empirical best linear unbiased prediction (Santner, Williams, and Notz 2003) with the MLEs of
the parameters is denoted by
\begin{linenomath}
	\begin{equation} \label{pred1}
	\Yhat (x_0) = f_0^t \; \ubhat + r_0^t \; {\hat V}^{-1} ( {\und y} - F
	\ubhat),
	\end{equation}\end{linenomath}
where $f_0$ is the known linear regression function vector, $F$ is a design matrix,
$r_0^t$ is the correlation vector between $Y(x_0)$ and model outputs $Y(X)$,
${\und y}$ is the vector of observations collected at the design sites, and $\hat \ubeta $ is the generalized least squares estimator of $\ubeta$
 (see the Supplemental Material for the details).

\par The combinations of $\beta$s and $\theta$s determine the
 model, but the following simple GP model is considered first:
 \begin{linenomath}
 \beq \label{model3}
 y(\ux)~=~ \beta_0 ~+~\beta_1 x_1 ~+~ ... ~+~ \beta_d x_d ~+~ Z(\ux) 
\eeq
\end{linenomath}
 with the correlation function (\ref{eqno2-2}) or (\ref{eqno2-3}). When a {\it common} $\theta$ of (\ref{eqno2-2}) is used, we call
 (\ref{model3}) ``Model 1" in this study. When the {\it several} $\theta_i$s of (\ref{eqno2-3}) are used, we call
 (\ref{model3}) ``Model 2".

\section{Methods for code tuning}
\label{anlse}

\subsection{Data structure}
For notational convenience, experimental data is denoted by the subscript ``E,"
computer simulation data by the subscript ``C," and ``both" computer and experimental data by ``B."
Let $\csta $ be an adjustable
parameter vector to be estimated. Let $\uc$ be the input variables of the computer code corresponding
to $\csta$. The original experimental input variables are denoted by $X$.
 Let $q$ and $p$ be the dimensions of $\csta $ and $X$.
Further, let $n_C and n_E$ be the number of observations; then, $n_B =n_C+n_E$.
The details of the data structure for calibration are described in the Supplemental Material.

\subsection{Approximate nonlinear least squares}

In this subsection, the GP model approach to be used in the
ANLS procedure provided in Cox, Park, and Singer (2001) is described. If the
parameters $\sigma^2$, $\ubeta$, and $\theta$ are known, then,
for a given value of ${\bf \tau}$, a ``prediction'' of ${\bf y}_E ({\bf \tau}, {\bf x}_{iE})$
can be calculated for the given computer data.
 This is obtained using equation (\ref{pred1}) and $X_E, ~F_E, ~y_E$, and $V_{EE}$.
 Here, the computer data alone are used to calculate $\ubhat$ and ${\hat \utheta}$, and the data are not used thereafter.
 Note that $X_E, F_E, R(X_E, X_C), R(X_E, X_E)$, and $\ubhat$ are functions of $\csta$.

A design site selected for a computer experiment is denoted by
 $(\uc, x_C )$. Then, the
computer response $y$ (or $y_C$) at $(\uc, x_C )$ is
\begin{linenomath}
\beq \label{yCCnew} y_C~=~Y (\uc, x_C ) , \eeq
\end{linenomath}
where $Y$ represents the expected value of the output from computer code. $Y$ can be viewed as the value produced from a theory.
Since we assume that the computer code is close to the real experimental data with some variation if the tuning parameters in the computer code are optimal, the response of the real experiment $y_E$ at $(\csta , ~x_E)$ is modeled by
\begin{linenomath}
\beq\label{yEE} y_E~=~ Y (\csta , ~x_E) ~+~ \epsilon_E.
 \eeq \end{linenomath}
 Here $Y (\csta , ~x_E)$ also represents the expected value of the response in the real experiment.
 This common $Y$ in the two abovementioned equations (\ref{yCCnew}) and (\ref{yEE})
 connects the computer code and the real experiment.
 The stochastic term $\epsilon_E$ is assumed to be independent and identically
distributed with mean zero and variance $\see2$.
On the other hand, a computer code is sometimes considered inexact because of computer model bias, which is the systematic difference between the model and the truth (Kennedy and O'Hagan 2001; Gramacy 2016; Plumlee 2017). In that case,
 the response of the real experiment can be modeled by
 \begin{linenomath}
\beq\label{yEbias} y_E~=~ Y (\csta , ~x_E) ~+~ b(\csta , ~x_E)  ~+~ \epsilon_E,
 \eeq \end{linenomath}
  where $b(\csta , ~x_E)$ represents the model bias or discrepancy.

\par
A drawback of the computer experiment is that one run of a complex simulator sometimes requires several minutes, say, $m$ minutes.
  Then, $\csta$ is usually estimated by minimizing the residual sum of the squares:
  \begin{linenomath}
\beq\label{RSS}
 RSS(\csta)~=~ \sum _{ i=1}^{ n_E} ~[~ {y_E}_i ~-~ Y (\csta,~ {x_E}_i )]^2,
\eeq \end{linenomath}
 where ${y_E}_i $ is an observed response from the real experiments, and
$Y (\csta,~{x_E} _i )$ is the expected value of the output
from the computer code at the experimental
point $(\csta,~ {x_E}_i )$. One evaluation of
$RSS(\csta)$ requires approximately $n_E \times m$ minutes. It is thus computationally infeasible
to run the code as many times as needed for an iterative nonlinear optimizer to find $\csta$.

\par
The ANLS method first fits the GP Model 1 or Model 2 as defined in equation (\ref{model3}), with MLE using computer data alone. Then, upon treating the fitted prediction model as if it were the
true model, the method attempts to determine the $\cshat$ that minimizes the residual sum of squares with predictors:
\begin{linenomath}
\begin{equation}\label{RSSp}
RSS_P ( \csta )~=~ \sum_{i=1}^ {n_E}~ [~{y_E}_i
-~ \Yhat ( \csta ,~ {x_E}_i ) ]^2,
\end{equation} \end{linenomath}
 where $\Yhat(\csta,~{x_E} _i)$ is the empirical best linear unbiased
 prediction of $Y(\csta,~{x_E} _i )$, as in (\ref{pred1}).
 Since it is difficult to have a closed-form minimizer of (\ref{RSSp}), a numerical optimization routine is necessary to determine
$\cshat$. Note that $\Yhat$ is a computationally cheaper emulator (surrogate or metamodel) of the
expensive simulation code. This makes the problem 
computationally feasible. It is applied for each GP model (Model 1 and Model 2) stated in (\ref{model3}).

\par The advantages of this method are that it is reasonably
 economical and easy to implement, and
the computer and experimental data are uncoupled.
 The ANLS method does not require the functional relation between the inputs and output to be the same for the computer code and real experiments. In contrast, the likelihood-based approaches described in the next subsection do require this, but they employ marginal likelihoods to estimate model parameters.
The prediction residuals, $r_i
= {y_E}_i - \Yhat(\cshat ,~{x_E} _i)$, can
 be used to check the validity of prediction model, $\cshat$,
and of the ANLS method. A potential drawback of the ANLS method is that it does
not account for uncertainty in the approximation of $Y_C$ by $\Yhat$. Another
difficulty is that the metamodel $\Yhat$ is built only once and is not
updated thereafter.

\subsection{Likelihood-based tuning methods}
  Given a computer code and our GP approach for $y$,
a unified statistical approach is available. We have the likelihood for all the
parameters, including the tuning parameters $\csta$; the error term parameter
$\see2$, and the random function parameters $\ubeta$, $\theta$,
and $\sigma^2$. Thus, all parameters can be estimated using the MLE method. We refer to this method as the full MLE. It can be done for each GP model defined in (\ref{model3}).
The $-2$
times concentrated log likelihood function (except for constants) of all parameters for the combined data with
$\hat\ubeta_B$ and $\hat{\sigma^2_B} $ plugged in is
\begin{linenomath}
\beq\label{constllk}
 -2\; log\; L(\csta, \eta_B; ~\uy_B,  X_B)~=~ n_B\; log\;
\hat{\sigma^2_B }~+~ log\; | V_B |, \eeq where \beq
 \hat{\sigma^2_B} = (\uy_B- F_B \ubhat_B ) ^t V^{-1}_B (\uy_B - F_B
  \ubhat_B ) / {n_B},
  \eeq
  \beq \label{betaB}
\ubhat_B ~= ~({F_B }^t {V^{-1}_B} F_B )^{-1} {F_B}^t {V^{-1}_B} \uy_B ,
  \eeq  \end{linenomath}
where $\eta_B = ( \utheta_B , \ubeta_B , \sigma^2_B , \ge2)$, where $\ge2 = \see2 / \sigma^2$.

Some other approaches based on the likelihood function, including the full MLE,
were also proposed by Cox, Park, and Singer (2001). One
of them is the ``Separated MLE" (SMLE) method,  which maximizes the
conditional likelihood function of experimental data when the computer data is given.
Here, the parameters $\ubeta$, $\utheta$, and $\sigma^2$ are
estimated by maximizing the marginal likelihood for the computer data only. These are plugged into the conditional likelihood of the
experimental data given the computer data.
 This is then maximized with respect to $\gamma_E$ and $\csta$ to obtain estimates of those
parameters. 

One advantage of the likelihood-based method is that it can simultaneously use both computer and
experimental data to estimate $\csta$, whereas ANLS method uses computer data
only. These likelihood-based approaches enrich the tuning methods.
 Cox, Park, and Singer (2001), found the SMLE method to be better than the full
MLE method. Thus, SMLE is compared with the proposed method in this study. The details of the SMLE are provided in the Supplemental Material.

\section{Proposed method}
\label{max-min}

\subsection{Max-min algorithm}
\label{subMn}
The following are the steps for the proposed tuning method.
We call it a Max-min algorithm because it uses maximization and minimization iteratively.
\\

{\bf Algorithm Max-min:} iterate Step 3 and Step 4 until convergence is achieved

\begin{itemize}
\item[] {\bf Step 1} (model building): build a surrogate (\ref{model3}) using the MLE for the given {\it computer data only}.

\item[] {\bf Step 2} (initial solution): set iteration $i=1$, and find $\cshat$ by minimizing $RSS_p (\csta)$ in (\ref{RSSp}) using surrogate (\ref{model3}).

\item[] {\bf Step 3} (maximization): build a new surrogate (\ref{model3}), using the MLE for the {\it combined data} with the fixed $\cshat$ obtained in the previous step.

\item[] {\bf Step 4} (minimization): set iteration $i=i+1$, and find $\cshat$ by minimizing $RSS_p (\csta)$ in (\ref{RSSp}) using the surrogate built in Step 3.
 If $\cshat$ satisfies the stopping rule, then stop; otherwise, go to Step 3.
\end{itemize}

Note that in each iteration of Steps 3 and 4, $\cshat$ is updated; thus, the estimates of the parameters of
$\theta, \ubeta, \see2$, and $\sigma^2$ are updated. We expect this to
positively influence the finding of $\cshat$ of Step 4. Steps 2 and 4 are the same in terms of minimizing $RSS_p (\csta)$, but Step 2 uses only computer data, while Step 4 uses either the combined or computer data.
Steps 1 and 3 are the same in terms of obtaining the MLE by maximizing the likelihood function, but Step 1 uses only computer data, while Step 3 uses the combined data.
The likelihood function in Step 3 is $L(\eta_B; ~\cshat, ~\uy_B, ~ X_B)$, which is used for the full MLE method. But here, the
tuning parameters are fixed as the $\cshat$ that was obtained in the previous step.
For the optimizations in Steps 2, 3, and 4, quasi-Newton numerical algorithms were used.

 On using the combined data in Step 3, we assume that the functional relation between the inputs and output is the same for the computer code and physical process.
 The computer data and the experimental data are linked by the use of this common response function, with $\cshat$ being the value for the tuning parameters in data from the physical process.
 
 The Max-min algorithm stops when one of the following rules is satisfied:  for $i=1,2,\cdots$,
 \begin{enumerate}
 	\item[1] (maximum iteration): Number of iterations reaches the pre-assigned maximum number,
 	\item[2] (minimum improvement): $RSS_p (\cshat_{i+1} ) > RSS_p (\cshat_i ) -ftol \;$ for `maxagain' consecutive iterations,
 	\item[3] (minimum relative improvement): $ { [{RSS_p (\cshat_{i+1} ) - RSS_p (\cshat_i )] } / {RSS_p (\cshat_i )} } > -ftol \;$ for `maxagain' consecutive iterations,
 \end{enumerate}
 where `ftol' is a pre-assigned small value for tolerance. Here $ RSS_p (\cshat_1)$ is the minimum value of
 $RSS_p$ obtained in Step 2, and $RSS_p (\cshat_i )$ is the minimum value obtained in Step 4 in the $i$-th iteration.

 When the $RSS_p$ in Step 4 is greater than that of Step 2 or that of the last iteration,
 a small random fluctuation on $\cshat$ is given before Step 3. Without this fluctuation setting, the algorithm stopped within four iterations. Based on our experience, the random fluctuation caused a reduction in $RSS_p$. However, it did not make the algorithm execute more than 20 iterations. 

When $x_0$ is a given site representing $(\cshat ,~{x_E} _i) $ in the experimental data ($X_E$),
the prediction formula needed for computing $RSS_p$ in Step 2 is    
\begin{linenomath}
\begin{equation} \label{predANLS}
\Yhat_C (x_0) = f_0^t \; \ubhat_C + r_{0C}^t \; {\hat V}^{-1}_{CC} ( {\und y}_C - F_C
\ubhat_C),
\end{equation} \end{linenomath}
where $f_0$ is the known linear regression function vector; $r_{0C}$ is the $n_C
\times 1$ correlation vector between $Y(x_0)$ and $Y(X_C)$, and $\ubhat_C =~(F_C^t
V^{-1}_{CC} F_C)^{-1} F_C^t V^{-1}_{CC} \;\uy_C$. The $\hat \theta_C$ is
plugged into $V_{CC}$, where $\hat \theta_C$ is the MLE from the computer data only in Step 1.

The prediction formula needed for computing $RSS_p$ in Step 4 is
\begin{linenomath}
\begin{equation} \label{predmmB}
\Yhat_B (x_0) = f_0^t \; \ubhat_B + r_{0B}^t \; {\hat V}^{-1}_B ( {\und y}_B - F_B
\ubhat_B),
\end{equation} \end{linenomath}
where $r_{0B}$ is the $(n_E + n_C) \times 1$ correlation vector between $Y(x_0)$ and  $Y(X_B)$. Here, $\hat
\theta_B$, $\ubhat_B$, and $\hat \gamma_E $ are the MLEs from the {\it combined data} in Step 3. 
Note that $\hat \theta_B$ and $\hat \gamma_E $ are needed for constructing $V_B$.
Even though $x_0$ and $X_E$ are included in the construction of $r_{0B}$ and $V_B$, the prediction 
$\Yhat (x_0)$ in (\ref{predmmB}) is not an exact interpolation because of the positive $\hat \gamma_E$.

Our approach was motivated by the iteratively re-weighted least squares method in regression analysis. The alternating estimation of parameters recursively is similar to the EM (expectation-maximization) algorithm. One can view this method as similar to a frequentist version of the Bayesian modularization approach (Liu, Bayarri, and Berger 2009). Some practical suggestions for the calibration of large-scale simulations and optimization are provided in Gramacy (2016, Section 4). However, our study was conducted independently from their study.

The advantages of the proposed Max-min algorithm are as follows. The
uncertainty in the approximation of $Y$ using $\Yhat$ in Step 4 is smaller than
that in the ordinary ANLS, because the MLE method in Step 3 and the prediction
model in Step 4 use combined data (a larger sample size), while the ordinary
ANLS method uses only the computer data in Step 1 for MLE and in
Step 2 for prediction. In addition, the Max-min algorithm accounts for the
covariance between the computer data and experimental data in building the
metamodel. Moreover, our test function experience revealed that the solutions
from the Max-min algorithm are less influenced by the initial value of $\cshat$
than by the ANLS method because the estimated $\cshat$ in the Max-min is iteratively
updated several times.

\subsection{Approximated confidence region of estimates}

Once $\csta$ has been estimated, some indications of the accuracy of the
estimates are generally necessary. Thus, we rely on the asymptotic
theory for the nonlinear least squares estimator that appears in the regression
analysis (Draper and Smith 1981).
The approximate
$100(1-\alpha) \%$ confidence region of $\csta$ is obtained by
\begin{linenomath}
 \beq \label{confr}
\{  \csta : RSS_P(\csta) \leq RSS_P(\cshat) [ 1 + {q \over {n_E -q }} F_\alpha (q,
n_E -q ) ] \},
\eeq \end{linenomath}
where $F_\alpha (q, n_E -q )$ is the upper $(1-\alpha)$
percentile of the $F$ distribution with $q$ and $n_E -q $ degrees of freedom ($q$
is the number of parameters in $\csta$). Wong, Storlie and Lee (2017) considered a Bootstrap approach to compute the uncertainty of the estimates.

\section{Toy-model study}

\subsection{Exact computer models}
In this section, we apply the methodology outlined in the previous
sections to the seven test functions $Y(\csta, x)$ that are
 easy to compute. In the first five test functions, the experimental data with
sample size $n_E$ are generated by
\begin{linenomath}
\beq \label{toyee} y_{E}= Y(\tau^*, x)+e, \eeq
\end{linenomath}
where the random variable $e$ follows a normal distribution with mean zero and
variance $\sigma_{e} ^{2}$. These five examples assume that the functional relation between the output and inputs is the same for both the computer code and the physical process.
By contrast, in the last two test functions (6 and 7), the model bias term $b(x)$ is added to the function
$Y(\tau^*, x)$ in (\ref{toyee}). 
Here, $\tau^{*}$ stands for the true tuning parameters
and is specified for each toy model. No random error is given to the computer
responses. Sample sizes for both data are set to
30 ($n_C = n_E =30$) for the first five test functions. For generating the
computer and experimental data, the same uniform
distributions were used for the random numbers of $x$ variables.
The details on the eight test functions are described in the Supplemental Material.

To address the uncertainty resulting from the designs for the input variables, we
repeated the $\csta$ estimation using 30 random Latin-hypercube designs. Thus, 30
sets of estimates of $\csta$ were obtained, and the averages and standard
deviations are reported.
Two different GP models of (\ref{model3}) were used with the correlation functions of (\ref{eqno2-2}) and (\ref{eqno2-3}), which are called ``Model 1" and ``Model 2," respectively.

\begin{figure}[tbh]
 	\begin{center}\includegraphics[width=13cm,height=9cm]{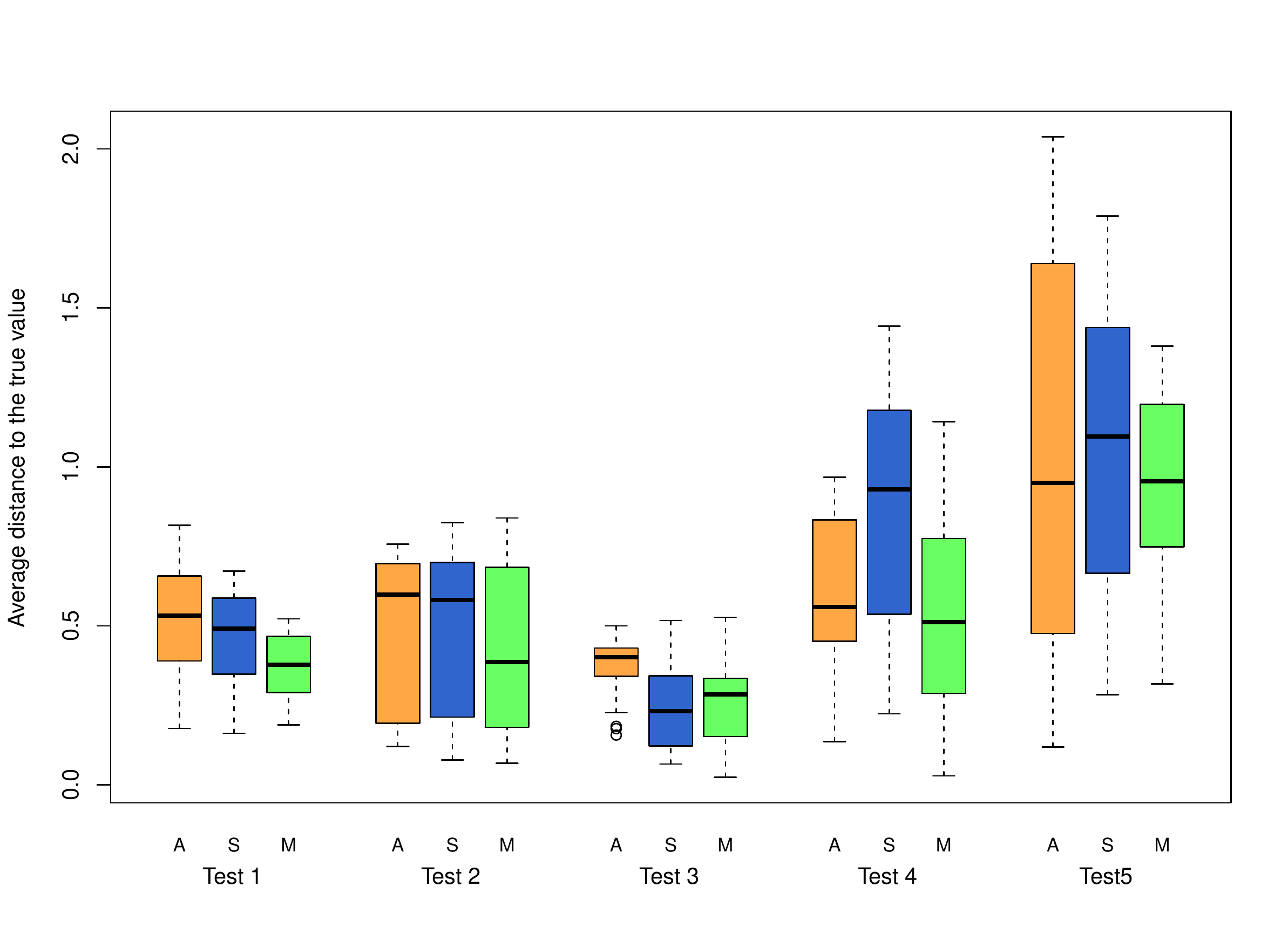}
 	\caption{ Box plot of distance to the true value in Five test functions in which Model 2 is employed as a surrogate. The acronyms A, S, and M at the bottom stand for the ANLS, SMLE, and Max-min methods, respectively.
 		\label{test_d_box}}\vspace{.2cm}
 	\end{center}
 \end{figure}

Figure \ref{test_d_box} shows box plots of the distance to the true value in five test functions in which Model 2 is employed as a surrogate. The Max-min algorithm generally works better than the ANLS and SMLE methods.
Tables S1 to S8 in the Supplemental Material present the results for each toy model,
each method, and Model 1 and Model 2. Some columns show the average of Euclidian
distances (Dist) between the true $\csta$  and estimates, with
standard deviations in parentheses. The last column shows the mean
squared error (MSE) of the estimates obtained using the following formula:
\begin{linenomath}
 \beq  MSE (\cshat) = (Dist)^2 +  \sum_{i=1}^q (std (\hat \tau_i))^2,
 \eeq \end{linenomath}
where $std (\hat \tau_i)$ is the standard deviation of each estimate computed from 30 repetitions.
Figures S1 to S5 show the box plots for each test function and method.
In terms of the average distance to the true value
and MSE in Tables S1 to S5, the Max-min algorithm works better than ANLS for all toy models.
Max-min also works better than SMLE
 for test functions 1, 2, and 4. Finally, SMLE works slightly better than ANLS, except in test function 2.

Figure S6 shows the typical convergence of the Max-min algorithm for
10 trials of test function 1. In most cases, the algorithm stops at the third or
fourth iteration, except for a few cases wherein it stops at the sixth or seventh
iteration. This means that the improvement of $\cshat$ in the second iteration is
significant, while it may not be
significant after the second iteration. 
The small random fluctuation on $\cshat$, as mentioned in subsection \ref{subMn}, was not applied in this computation.

\subsection{Another prediction}
A potential drawback of the Max-min algorithm is that it partially fails to decouple between the estimation of $\csta$
and building of the metamodel because it uses the combined data in both Steps 3
and 4. To address this difficulty, instead of using equation (\ref{predmmB}) in Step 4, one can employ prediction (\ref{predANLS}) but with
$\hat \theta_B$ and $\ubhat_B$ (or $\ubhat_C$): 
\begin{linenomath}
\begin{equation} \label{predCgB}
\Yhat (x_0) = f_0^t \; \ubhat_B + r_{0C}^t \; {\hat V}^{-1}_{CC} ( {\und y}_C - F_C
\ubhat_B).
\end{equation} \end{linenomath}
This leads to another version of the Max-min algorithm, which is applied to the next toy models. 
Note that this version does not use $\hat \gamma_E$. We denote this prediction as $\Yhat_{C|B} (x_0)$.
We can employ $\ubhat_C$ instead of $\ubhat_B$ in (\ref{predmmB}), but we have not yet attempted to do so.

\subsection{Inexact computer models}

A computer model is often considered inexact. This means that the computer model does not perfectly match the real system even if some parameters included with it are optimal (Plumlee 2017). This is the computer model bias; it is the difference between the model and the truth.
The model-bias term $b(x)$ is included in the following two test functions.
These are employed to check the performance of the proposed method for cases where the functions generating the output for the computer code and for the physical experiment differ, but we incorrectly assume that
(\ref{yCCnew}) and (\ref{yEE}) hold.

\vspace{.2 cm} Test function 6:  $n_C = n_E =20$
\begin{linenomath}
	\begin{eqnarray*}
& Y(\csta,x) =\tau(1) x_1^2 + \tau(2) x_2 \\
& Computer ~~ data: T_1 \sim U(1,8), ,\;  T_{2} \sim U(1,8), \;
 x_{1} \sim U(0, 1),\;  x_{2} \sim U(0, 1)\\
& Experimental~~ data: \; y_E =Y (\csta , ~x) +b(x)+\varepsilon ,~
        b(x) =x_2 sin(5 x_2), \; \\
&  \tau_{1} =4.0 ,\;  \tau_{2} = 4.0,\;
   \sigma_{E}^{2} =0.02^2.
	\end{eqnarray*}
\end{linenomath}

 \vspace{.2 cm} Test function 7: $n_C = n_E =20$
\begin{linenomath}
	\begin{eqnarray*}
&Y(\csta ,x)=
	 \Big(1-exp(-\frac{1}{2x_2})\Big) \times
	 {(100 \tau_1 x^3_1+1900x^2_1+2092x_1+60)}/{(100  \tau_2
        x^3_1+500 x^2_1+4x_1+20)} \\
         & +5\; exp(-\tau_1) \times ({{x_1}^{{\tau_3}/10}})/({100({x_2}^{2+({\tau_3}/10)} +1)})\\
  &Computer ~~ data: T_{1}, \;T_2, \;T_3 \sim U(0.1, 5) ,\;  
   x_{1},\;   x_{2} \sim U(0, 1)\\
  & Experimental~~ data: \; y_E =Y (\csta , ~x) +b(x)+\varepsilon ,~
        b(x) =({10x^2_1+4x^2_2 })/({50x_1x_2+10}), \; \\
  & \tau_{1} =2.0 ,\;  \tau_{2} = 1.0,\;  \tau_{3} = 3.0,\;
    \sigma_{E}^{2} =0.5^2.
	\end{eqnarray*}
\end{linenomath}

 Test function 6 is modified from Plumlee (2017).
 Test function 7 was used in Bastos and O'Hagan (2009), in Goh et al.(2013), and in Gramacy (2016). 
 Twenty random Latin hypercube designs were used repeatedly to address the uncertainty.
 
In the previous exact toy-model study, we used the difference between $\cshat$ and true $\csta$ 
 for the physical process as a measure of performance. This requires that there be a true $\csta$,
 but there would be no such true $\csta$ in an inexact computer model.
 In this situation, the minimum value of $RSS_p$ would be a more appropriate measure.
 
   In calculating $RSS_p$ in the Max-min algorithm for the above two test funtions, we set ftol = 1.e-4, and maxagain = 7. For the small fluctuation of $\cshat$, random numbers from $N(0, \sigma_\tau^2)$ were used where $\sigma_\tau=max(\cshat\times 0.1,\; 0.3)$.
   
\begin{figure}[!tbh]
   	\centerline{\includegraphics[width=12cm,height=7cm]{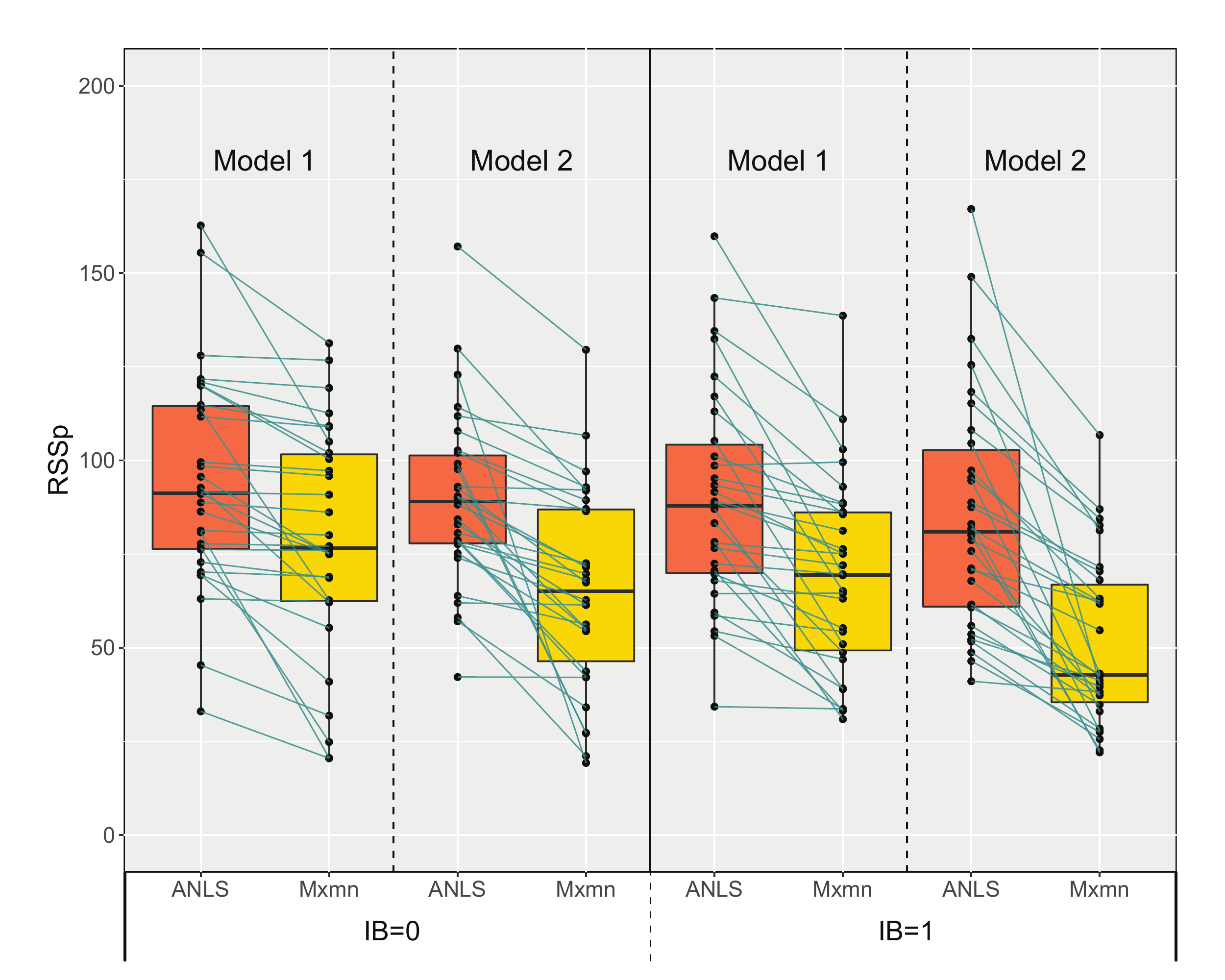}}
   	\caption{ $RSS_p$ (residual sum of squares with prediction) values for test function 6 computed from 20 Latin hypercube designs. $RSS_p$ values were calculated via two tuning methods with GP Model 1 and Model 2, and via two different predictions.
   		The acronyms IB=0 and IB=1 stand for prediction by $\Yhat_{C|B}$ in (\ref{predCgB}) and by  $\Yhat_B$ 
   		in (\ref{predmmB}) in the Max-min algorithm, respectively.  	   		\label{test6_box}}
   \end{figure}

\begin{figure}[!tbh]
   	\begin{tabular}{c}
   		\includegraphics[width=7cm, height=6cm]{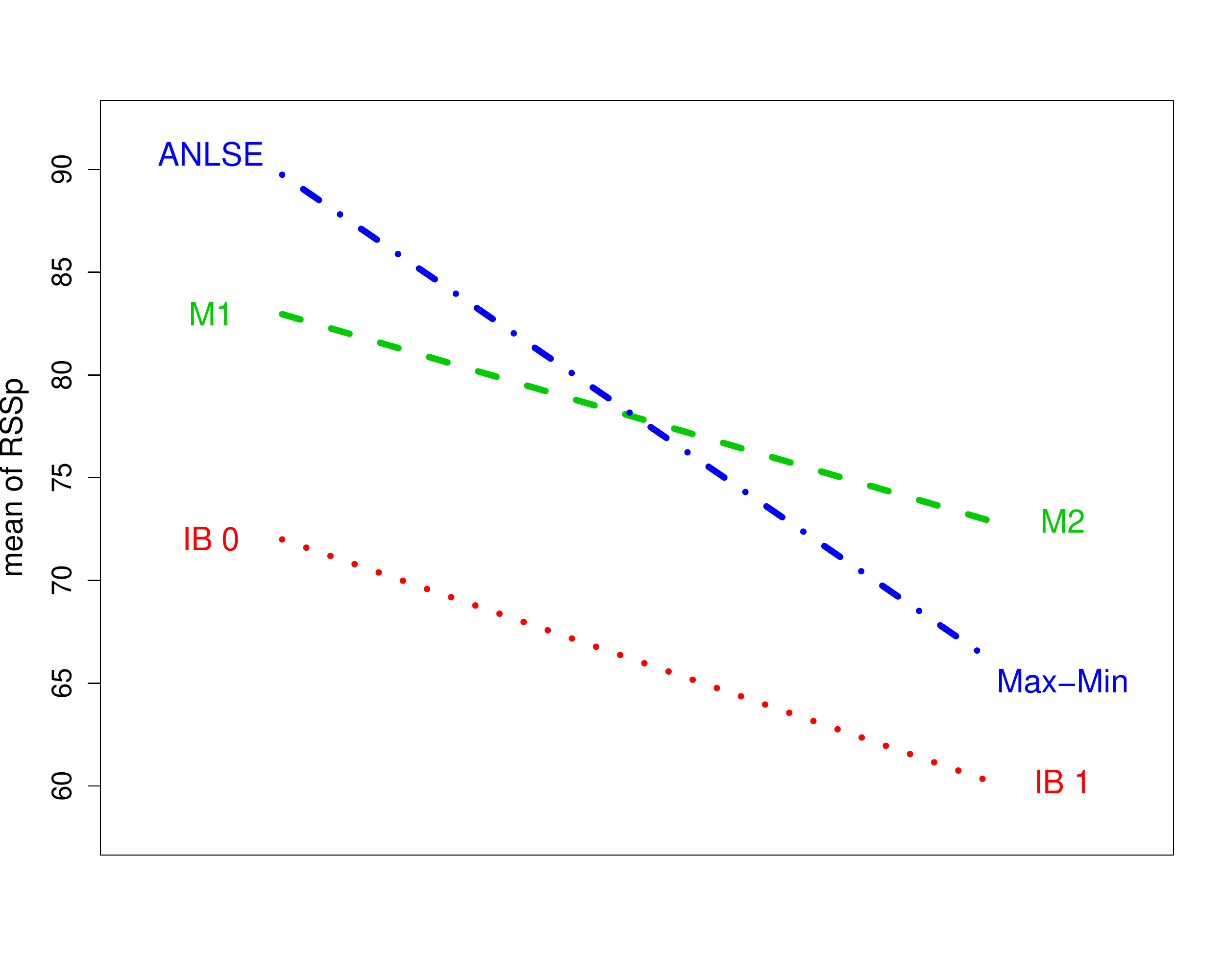}
   		\includegraphics[width=7cm, height=6cm]{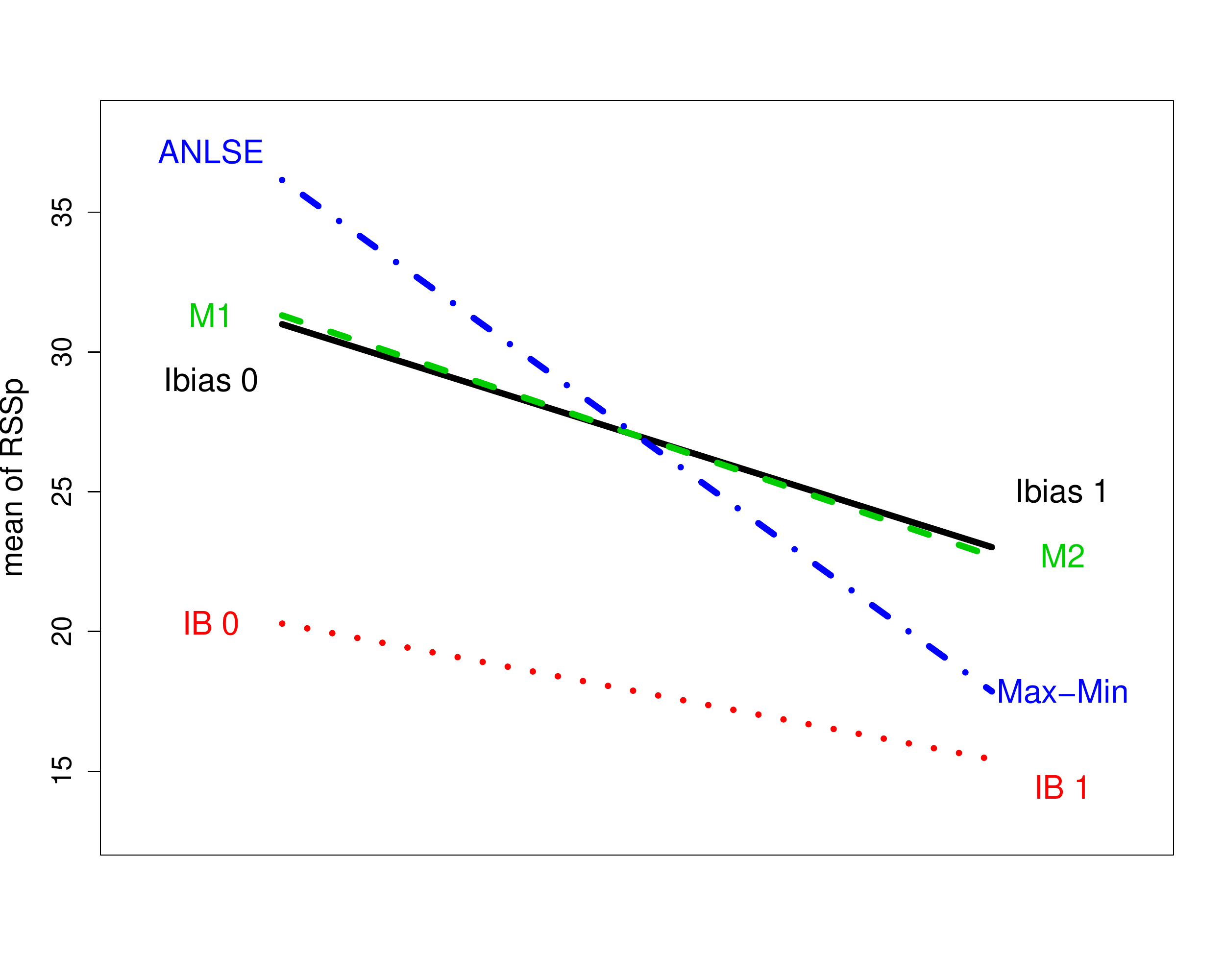}
   	\end{tabular}
   	\caption{ Mean values of $RSS_p$ (residual sum of squares with prediction) for test function 6 (left panel) and for test function 7 (right panel), showing the tuning methods (ANLS and Max-min), surrogate models (M1 and M2), and prediction formulas (IB0 and IB1). 
   		The acronyms M1 and M2 represent for GP Model 1 and Model 2, IB0 and IB1 for the predictions by $\Yhat_{C|B}$ and $\Yhat_B$ in the Max-min algorithm, and `ibias 1' and `ibias 0' for the cases with and without bias correction, respectively.  		
   		\label{test67_maineff}}\vspace{.2cm}
   \end{figure}

   Figure \ref{test6_box} shows parallel coordinated box plots of $RSS_p$ values for test function 6 computed from 20 Latin hypercube designs. $RSS_p$ values are calculated using the ANLS method and the Max-min algorithm with GP Model 1 and Model 2 via two different predictions. The predictions by $\Yhat_B$ in (\ref{predmmB}) and $\Yhat_{C|B}$ in (\ref{predCgB}) in calculating $RSS_p$ in the Max-min algorithm were attempted for the sake of comparison. The Max-min algorithm worked better than ANLS in most cases. The left panel of Figure \ref{test67_maineff} is an ANOVA-type main-effect plot based on mean values, which shows 
   	improvements from ANLS to Max-min, from Model 1 to Model 2, and from $\Yhat_{C|B}$ to $\Yhat_B$. The mean values from the last two predictions (IB0 and IB1) are calculated within the Max-min algorithm.

   	Relative improvement (RI) from ANLS to Max-min is provided in Table \ref{tab_test67}. RI is defined by
   	\begin{linenomath}
   		\begin{equation} \label{RI}
   		RI = { {{mean ~ RSS_p (ANLS)}\;-\;{mean ~ RSS_p (Max-min)}} \over {mean ~ RSS_p (ANLS)} } \time 100.
   		\end{equation} \end{linenomath}

   \begin{table} [!tbh]
   	\centering
   	\caption{Relative improvement in \% defined in (\ref{RI}) from ANLS to the Max-min method for test functions 6 and 7.
   		The acronyms `ibias 1' and `ibias 0' stand for the cases with and without bias correction, and IB = 0 and IB = 1 represent the predictions by $\Yhat_{C|B}$ and $\Yhat_B$ in the Max-min algorithm, respectively. }
   	\label{tab_test67}
   	\begin{center}
   		\begin{tabular}{c|cc|cc|cc}
   			\hline
   			\multirow{2}{*}{} & \multicolumn{2}{c|}{Test 6 ibias 0}   & \multicolumn{2}{c|}{Test 7 ibias 0 } & \multicolumn{2}{c}{Test 7 ibias 1}\\
   			IB &  \multicolumn{1}{c}{Model 1} & \multicolumn{1}{c|}{Model 2} & \multicolumn{1}{c}{Model 1} & \multicolumn{1}{c|}{Model 2} & \multicolumn{1}{c}{Model 1} & \multicolumn{1}{c}{Model 2}  \\ \hline
   			0 ($\Yhat_{C|B}$) & 16.5 & 27.0 &24.5 & 61.4 & 23.0 & 69.5  \\
   			1 ($\Yhat_B$) & 23.1 & 39.9 & 52.4 & 37.4 & 76.2 & 74.6 \\ 	\hline
   		\end{tabular}
   	\end{center}
   \end{table}

\begin{figure} [!tbh]
	\begin{tabular}{c}
		\includegraphics[width=15cm, height=8cm]{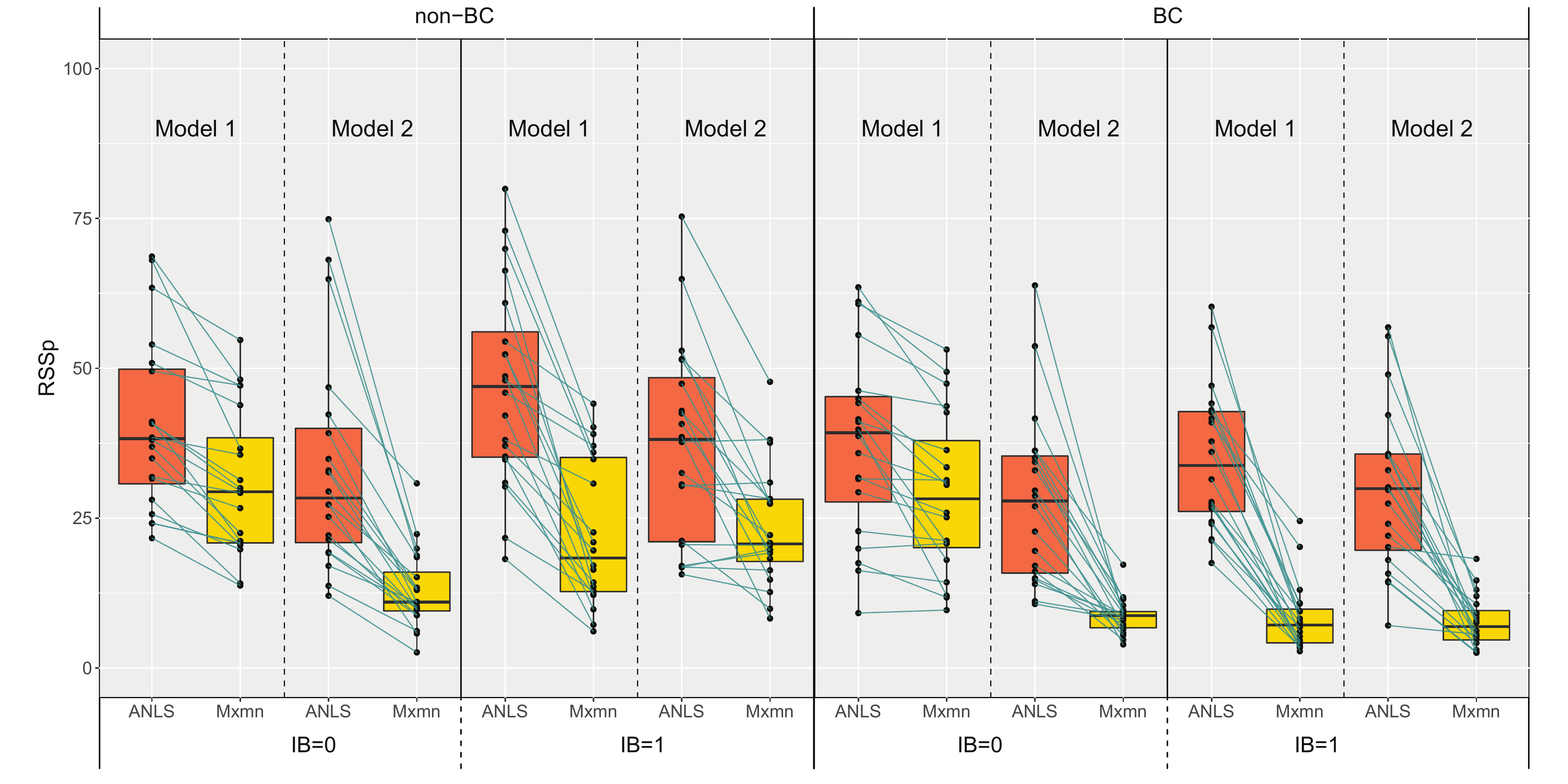}
	\end{tabular}
	\caption{This is the same as Figure \ref{test6_box}, but for test function 7,  calculated without (non-BC) and with bias correction (BC).
		\label{test7_box}}
\end{figure}

For tuning test function 7, we considered a simple bias correction (BC) technique. Two constants for additive and multiplicative corrections were set for the predictor (Fernandez-Godino et al. 2016) as
 \begin{linenomath}
 	\begin{equation} \label{predBC}
 	\Yhat_{bc} (x_0) = \rho\; \Yhat (x_0) +\delta.
 	\end{equation} \end{linenomath}
 This naive BC method may not be better than sophisticated methods such as that using $\rho(x)$ and $\delta(x)$, or the approach by Kennedy and O'Hagan (2001), but we hope for this to work better than the non-BC case.
 The objective function $RSS_p$ in (\ref{RSSp}) is now changed to 
 \begin{linenomath}
 	\begin{equation}\label{RSSpbc}
 	RSS_p^{bc} ( \csta, \rho, \delta )~=~ \sum_{i=1}^ {n_E}~ [~{y_E}_i
 	-~ (\rho \;\Yhat ( \csta ,{x_E}_i ) + \delta ) ]^2.
 	\end{equation} \end{linenomath}
Then, $\csta$, $\rho$, and $\delta$ are estimated simultaneously by minimizing $RSS_p^{bc}$ in both the ANLS and the Max-min methods.

Figure \ref{test7_box} shows parallel coordinated box plots of the $RSS_p$ values for test function 7, computed from 20 Latin hypercube designs. The left panel is for non-BC, while the right one is for BC. 
The acronyms in this figure are the same as in Figure \ref{test6_box}. This figure shows that the Max-min algorithm worked better than ANLS in most cases. The right panel of Figure \ref{test67_maineff} is a main-effect plot for test function 7. It shows 
improvements from ANLS to Max-min, Model 1 to Model 2, non-BC to BC, and $\Yhat_{C|B}$ to $\Yhat_B$.
Note that the patterns of the main effects for test functions 6 and 7 are very similar.
The RI from ANLS to Max-min is provided in Table \ref{tab_test67}. The bias correction was most effective when used with Model 2 and with the prediction $\Yhat_B$.

\section{Application to nuclear fusion model}

\subsection{Nuclear fusion data}

A simple measure of energy efficiency in a nuclear fusion device (called a tokamak, from the Russian language) is the
global energy confinement time $\tau_E$. The theoretically based
confinement model can be written as follows (Kay and Goldston 1985):
\begin{linenomath}
\beq   y_ E ~=~ f(~ \csta ,~ P,~I,~N,~B) ~,
\eeq \end{linenomath}
where $f$ is a known function calculated using a complex
simulation code called Baldur, $P$ is the total input power, $I$
is the plasma current, $N$ is the electron density, $B$ is the
magnetic field, and
 $ \csta ~=~(~\lcone ~ ,\lctwo,~\lcthree,~ \lcfour ~)$
 are the following adjustable parameters that determine energy transfer, that is, drift waves, rippling,
resistive ballooning, and the critical value of $ \eta_i $ (which provokes increased ion energy losses for the drift waves), respectively.

The experimental data comprises only $P,~I,~N,~B$, and the real
observation $y_E$, whereas the computer data comprises eight independent
variables $(\uc ,~ P,~I,~N,~B)$ and computer response $y_C$ obtained using
Baldur. The experimental data were drawn from the database of S. Kaye: 42
observations from the PDX (Poloidal Divertor Experiment) tokamak in Princeton
and 64 from the Baldur simulator (Singer et al. 1988). Because the Baldur simulator
requires five CPU minutes on a Cray supercomputer for one execution, a careful
selection of input points is required, which is a statistical design problem for a
complex simulation code. For this purpose, we used a data-adaptive sequential optimal design
strategy, which is described briefly in the Supplemental Material.

\begin{table}[!tbh]
	\centering
	\caption{Tuning results from nuclear fusion example where $\hat{\tau}$ are estimates of tuning parameters. $RSS_p$ is the residual sum of squares with predictor in which a GP Model 1 is employed.}     \vspace{.3 cm}
	\begin{tabular}{cC{1.5cm}C{1.5cm}C{1.5cm}C{1.5cm}C{1.5cm}}
		\hline
		Method &  $\hat{\tau_{1}}$  & $\hat{\tau_{2}}$ & $\hat{\tau_{3}}$ & $\hat{\tau_{4}}$ & $RSS_p$\\
		\hline
		ANLS  & 1.012  & 2.035  & 1.110  & 1.308  & 0.4406 \\
		SMLE  & 1.120  & 2.055  & 0.118  & 1.303  & 0.2908 \\
		Max-min  & 0.667  & 1.053  & 0.477  & 1.823  & 0.1546 \\
		\hline
	\end{tabular}%
	\label{tab_tok}%
\end{table}%

\subsection{Estimation and analysis}

Table S9 in the Supplemental Material presents the MLE of the parameters of GP Model 1 obtained from the combined data.
Table \ref{tab_tok} provides the results of $\csta$ estimation obtained using ANLS, SMLE, and the Max-min methods on the basis of Model 1.
 A quasi-Newton optimization routine was employed in searching for $\cshat$.
Several starting values were tried to avoid the local minima.
The last column in Table \ref{tab_tok} shows the value of $RSS_p$ at the convergence of the
algorithms. The $RSS_p$ for SMLE was obtained by calculating $RSS_p$ for the
estimated $\csta$. The $RSS_p$ for the Max-min with Model 1 is the smallest among the 
three methods.

\begin{figure}[tbh]
	\centerline{\includegraphics[width=12cm,height=9cm]{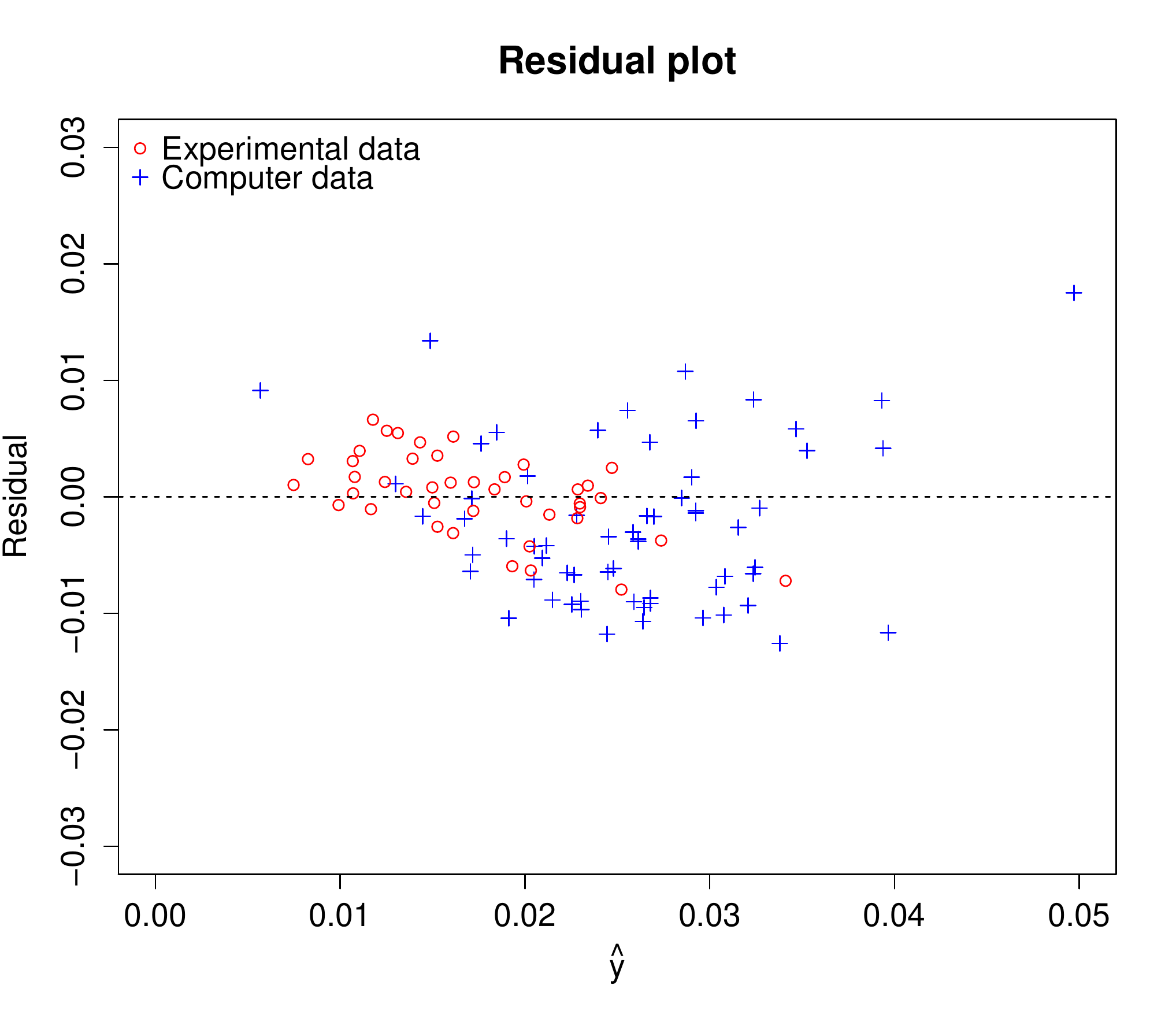}}
	\caption{Nuclear fusion example in which the computer code was tuned by the Max-min algorithm using GP Model 1. The circles and crosses denote experimental and computer data, respectively.
		\label{tok_res}
	}
	\vspace{.2cm}
\end{figure}

\begin{figure}[tbh]
	\centerline{\includegraphics[width=12cm,height=9cm]{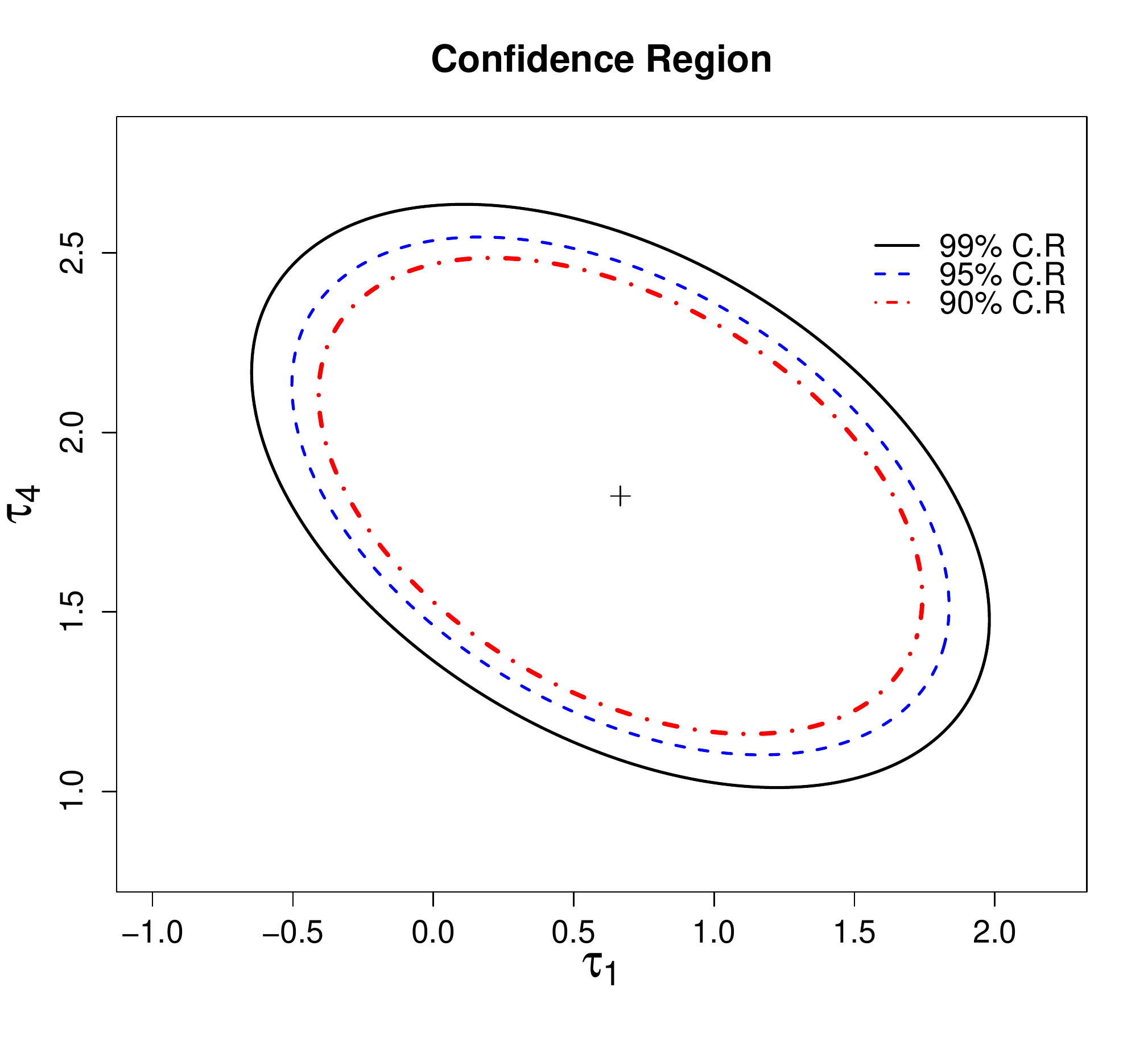}}
	\caption{Tuning parameters in Baldur code for nuclear fusion data, where the estimate $(\hat \tau_1,`\;\hat \tau_4) = (.667,\; 1.823)$ is the center point. This was obtained using the Max-min algorithm using GP Model 1.
		\label{tok_cr}
	}
	\vspace{.2cm}
\end{figure}

Figures \ref{tok_res} and \ref{tok_cr} show the residual plot (residual vs. predicted values) and confidence regions of the tuning parameters ($\tau_1$ and $\tau_4$) for the tokamak
data, which were obtained using the Max-min algorithm with Model 1.
In Figure \ref{tok_res}, the predicted values for the computer data (circles) had
a wider horizontal range than those for the experimental predicted values (crosses). In addition, the residuals for the computer data had a narrower
vertical range than those for the experimental data. These results indicate that 
the computer data are fitted relatively well, and that good coverage of the
range of experimental observations is ensured.

\section{Summary and Discussion}

Using a GP model, we considered an iteratively re-estimated ANLS method to tune complex computer code to data,
 namely a Max-min algorithm. This method is an extension of the ANLS method.
 A simulation study using toy functions suggests that the proposed Max-min algorithm works better than the ANLS and SMLE methods.
 We applied this technique to a computational nuclear fusion model. The proposed method can be useful for other applications in disciplines in which unknown theoretical parameters
must be estimated using complex computer codes and real experimental data.
Moreover, the Max-min algorithm is also applicable to other metamodels, such as spline (Wong, Storlie and Lee 2017), support vector machines, and neural networks.

The Max-min algorithm requires more computing time than the ANLS method. This is because Max-min uses
the combined data, resulting in a larger correlation matrix that must be inverted. It also requires more iterative minimizations of $RSS_p$ compared to the ANLS method, which requires just one iteration.

Two versions of predictions were tried in the last two toy models wherein bias functions were added. The prediction based on both data sources ($\Yhat_B$) reduced $RSS_p$ (residual sum of squares of prediction) more than that based on computer data given the parameter estimates obtained from both data sources ($\Yhat_{C|B}$). Even though $\Yhat_B$ reduced $RSS_p$  more than $\Yhat_{C|B}$, the improvement of the estimation of true $\csta$ compared to $\Yhat_{C|B}$ is not known.
A simple bias correction was considered in test function 7. This bias correction was the most effective when used with GP Model 2 and with the prediction based on both data sources.

We tried two different GP models in the toy-model study,
and found that the estimates of tuning parameters were changed according to the selected GP model. Model 1 in (\ref{model3}) is a superimposition of a GP with correlation function (\ref{eqno2-2}) on the first-order regression model. We believe that the first-order regression part sometimes works well to fit the dominant relationship, even though the isotropic correlation (\ref{eqno2-2}) may be unrealistic for capturing the variability of responses from the dominant relationship. It is computationally easier to obtain the MLE when the number of input variables is large. Model 2, with (\ref{eqno2-3}) in (\ref{model3}) assumes a separable Gaussian correlation, which is a popular family of correlation models as per the literature (Santner, Williams and Notz 2018), but it is
computationally expensive to obtain the MLE if many inputs exist.

A model selection procedure among many GP models with various combinations of
nonzero $\beta_j$s and $\theta_j$s in (\ref{model3}), as in Marrel et al. (2008), Chen et al. (2016), and Lee and Park (2017), may lead to realization of a better surrogate model. In this study, we did not consider the identifiability of the tuning parameter, which is an important focus of recent developments (e.g., Plumlee 2017; Tuo and Wu 2018).

There are, however, basic limitations with tuning computer
code to real-world data regarding the experimental design. We found that the performance of tuning methods is significantly dependent
on the designs for both the computer experiments and the physical experiments. Some authors, including Cailliez, Bourasseau, and Pernot (2014), and Beck and Guillas (2016), explored this topic. We agree that a sequential tuning approach is practically useful, as in Pratola et al. (2013), Kumar (2015), and Damblin et al. (2018).
Further research on relevant designs under the sequential tuning approach will be helpful.

\renewcommand{\baselinestretch}{1.0}
\small
\section*{Acknowledgments}
 The authors would like to thank the reviewers and the associate editor for their helpful suggestions, which have greatly improved the presentation of this paper.
This report follows some parts of an invited presentation by the third author at the GdR
Mascot-Num annual conference at Ecole de Mines St-Etienne, France, in 2015. We
would like to thank the organizers of this conference for the invitation and for their
hospitality.
We are also grateful to Professor Clifford Singer (Department
Of the Nuclear Engineering department, University of Illinois at Urbana-Champaign) for providing the
tokamak data.
This work was supported by the National Research Foundation of Korea (NRF)
grant funded by the Korean government (MSIP) (no. 2016R1A2B4014518).
 Seo's work was funded by the Korea Meteorological Administration Research and Development Program
  ``Enhancement of Convergence Technology of Analysis and Forecast on Severe Weather" under Grant 1365003081.


\newpage
\normalsize


\section*{Supplemental Material}

\subsection*{A.1 Maximum likelihood estimation in Gaussian process model}

Once the data have been collecte at the observation sites $\{\underline x_1,...,
\underline x_n \}$, we use the maximum likelihood estimation (MLE) method to
estimate the parameters in linear model part and covariance function. Since we assume that
$y(x)$ is a Gaussian process with mean $F \ubeta $ and
covariance matrix $\sz2 ~R$, the likelihood function of $\uy$ is
\begin{linenomath}
	\beq \label{like}
	L(\uy;~ \utheta , \ubeta , \sz2 , \gc2 ,~ \ux)~=~ { (2 \pi
		\sz2 )^{ -n/2}
		\over  \sqrt{|~V|} }
	~ exp \left (~ -~ { (\uy-F \ubeta ) ^t  V^ {-1} (\uy-F
		\ubeta )
		\over 2~ \sz2 }~ \right ) ~,
	\eeq
\end{linenomath}
where $F$ is a so-called design matrix.
When the covariance parameters
$\utheta$ and $\gc2$ are specified, the MLEs of $\sigz2$ and $\ubeta$ are denoted by
\begin{linenomath}
	\beq \label{sz2beta}
	\ubhat ~=~(F^t V^{-1} F)^{-1} F^t V^{-1} \uy, ~~~~~~
	\szhat ~=~{ 1 \over n} (\uy-F \ubhat )
	^t  V^ {-1} (\uy-F \ubhat ).
	\eeq
\end{linenomath}
Here, $\hat \ubeta $ is the generalized least squares estimator of $\ubeta$.
Since the likelihood equations do not lead to a closed-form solution, a numerical
optimization procedure is required. The Cholesky decomposition $V = U^t U$ is used as a major computation in calculating the likelihood function, where $U$ is
an upper triangular Cholesky factor. The computational details of calculating and
minimizing negative log-likelihood function are provided in Park and Baek (2001).  One
can use a R program {\it DiceKrig} (Roustant 2012).

\subsection*{A.2 Data structure for code tuning}
\label{data-structure}

\subsubsection*{A.2.1 Computer and experimental data}

For notational convenience, experimental data is denoted by the subscript ``E"
and computer simulation data by subscript ``C." Let $\csta $ be an adjustable
parameter vector to be estimated. Let $\uc$ be the input variables of the computer code corresponding
to $\csta$. Here, $\csta$ is a vector of the deterministic tuning parameters and
$\uc$ is a vector of the random variables.

The original experimental input variables is denoted by $X$.
Let $q$ and $p$ be the dimensions of $\csta $ and $X$.
Further, let $n_C, n_E$ be the number of observations. Then, we have
the data matrix of the independent variables: $X_C$ and $X_E$ for
computer and experimental data;
\begin{linenomath}
	\beq \label{XE}
	X_E  \;= \; \left [
	\begin{matrix}
		\tau_1 & \tau_2 & \cdots & \tau_q & x_{E11} & x_{E21} & \cdots & x_{Ep1} \cr
		\tau_1 & \tau_2 & \cdots & \tau_q & x_{E12} & x_{E22} & \cdots & x_{Ep2} \cr
		\vdots & & & \vdots & & & & \vdots \cr
		\tau_1 & \tau_2 & \cdots & \tau_q & x_{E1n_{E}} & x_{E2n_{E}} & \cdots & x_{Epn_{E}} \cr
	\end{matrix}
	\right ]
	\eeq
	\beq \label{XC}
	X_C  \;= \; \left [
	\begin{matrix}
		t_{11} & t_{21} & \cdots & t_{q1} & x_{C11} & x_{C21} & \cdots & x_{Cp1} \cr
		t_{12} & t_{22} & \cdots & t_{q2} & x_{C12} & x_{C22} & \cdots & x_{Cp2} \cr
		\vdots & & & \vdots & & & & \vdots \cr
		t_{1n_{C}} & t_{2n_{C}} & \cdots & t_{qn_{C}} & x_{C1n_{C}} & x_{C2n_{C}} & \cdots & x_{Cpn_{C}} \cr
	\end{matrix}
	\right ].
	\eeq \end{linenomath}
\vskip8pt \noindent Here, $t_{ij}$ in $X_C$ represents the $j$-th value of the $i$-th $T$ variable ($T_i$) and $x_{Eij}$ and $x_{Cij}$ denote
the $j$-th value of the $i$-th $X$ variable of experimental ($X_{Ei}$) and computer ($X_{Ci}$) input.
$X_E$ is a $n_E \times (q + p)$ matrix and $X_C$ is a $n_C \times (q + p)$ matrix.
Note that the first part of $X_E$ is composed of the unknown parameters $\tau_1, \cdots,\tau_q$, while the corresponding part of
$X_C$ comprises input values ($t_{ij}$).

\subsubsection*{A.2.2 Combined data}

The following notations for combined computer and experimental data are introduced:
\begin{linenomath}
	\beq \label{XFy}
	X_B = \left ( {X_C \atop X_E}  \right ), ~~~~~
	F_B = \left ( {F_C \atop F_E } \right ) = \left ( {f(X_C) \atop f(X_E)} \right ), ~~~~~
	\uy_B= \left ( {\uy_{C} \atop \uy_{E}} \right )
	\eeq \end{linenomath}
for the data matrix of the independent variables; the so-called ``design matrix,” defined
as the functions of the values of input variables; and the computer responses and
real observations, respectively. Here, the subscript ``B" indicates the combined ``both" computer and experimental data. Note that variables $\uc$ are
incorporated in the simulation code as design sites. $X_C$ and $F_C$
contain $\uc$, while $X_E$ and $F_E$ are the functions of the unknown parameters
$\csta$.

The Gaussian process model is now simultaneously applied to computer and
experimental data. Let $\eta = (\csta, \utheta, \gc2, \ge2, \sz2, \ubeta)$, where
$\gc2=\sec2 / \sz2 $ and $\ge2=\see2 / \sz2 $, which are the variance ratios for
the computer and experimental data. Here, $\sec2$ and $\see2$ are the
variances of error term ($\epsilon$) in the Gaussian process model for the computer and
experimental data, respectively. When necessary, ${\ubeta}_C$ and
${\ubeta}_E$ are used to denote the regression coefficients for the computer and
real experimental data. Then, given the independence and normality
assumptions, we have
\begin{linenomath}
	\beq \label{lawy} Law ~( \uy_B | \eta ) = N ( F_B {\ubeta}_B ,~  V_B ),
	\eeq \end{linenomath}
where
\begin{linenomath}
	\beq \ubeta_B=  ( {\ubeta_{C},\; \ubeta_{E}}  )^t, \eeq
	\begin{equation} \label{matv}
	V_B \;=\left[ \begin{array}{cc}
	V_{CC}  &
	V_{CE} \\
	V_{EC}  &
	V_{EE}
	\end{array} \right]
	=
	\; \sigma^2 \left [
	\begin{matrix}
	R(X_C, X_C) & R(X_C, X_E) \cr R(X_E, X_C) & R(X_E, X_E)
	\end{matrix}
	\right ] ~+~ \sigma^2 \left [ 
	\begin{matrix}
	\gc2 I & 0 \cr 0 & \ge2 I
	\end{matrix}
	\right ],
	\end{equation} \end{linenomath}
\vskip8pt \noindent where $R(X_C, X_E)$ represents a $n_C \times n_E$ matrix
composed of the correlations computed between $X_C$ and $X_E$. Note that
$V_B$ is a $n_B  \times n_B$ positive definite covariance matrix for the
combined data, where $n_B =n_C+n_E$. We set $\gc2 =0$ because only  a deterministic computer model is considered in this study.

\subsection*{A.3 Separated MLE for code tuning}

For the details of SMLE, we  make use of the conditional distribution of the experimental
data given the computer data, which is normally distributed with mean
\begin{linenomath}
	\begin{equation} \label{meanE|C}
	\mu_{E|C} \speq
	E[ {\uy}_E | {\uy}_C ;~ \csta, \eta ]
	\speq
	F_E \ubeta_E + \spp V_{CE}^t \;V^{-1}_{CC} \; ({\uy}_C -  F_C \ubeta_C ),
	\end{equation}
	and covariance
	\begin{equation} \label{covE|C}
	V_{E|C} \speq
	\mbox{Cov} [ {\uy}_E | {\uy}_C ; ~\csta, \eta]
	\speq
	V_{EE} \spm V_{CE}^t\; V^{-1}_{CC} \;V_{CE} ,
	\end{equation}  \end{linenomath}
where covariance matrices $V$'s are given as in (\ref{matv}).
In these formulae, we suppressed  the parameter dependencies in
$\mu_E= F_E \ubeta_E$, $\mu_C=F_C \ubeta_C$, $V_{CE}$, $V_{CC}$, and $V_{EE}$. Now the $-2$ times
concentrated log conditional likelihood function (except for constants) with
$\hat\ubeta$ and ${\hat\sigma^2}_{E|C}$ plugged in is
\begin{linenomath}
	\beq\label{llkeSMLE}
	-2\; log\; L(\csta, \gamma_E; X_E, \uy_E |\; \uy_C,
	{\hat\gamma}_C, \ubhat_C, \hat\utheta,  X_C)~=~ n_E\; log\;
	{\hat\sigma^2}_{E|C} ~+~ log\; | V_{E|C} |, \eeq where
	\beq
	{\hat\sigma^2}_{E|C} = (\uy_E- {\hat\mu}_{E|C} )^t \; V^{-1}_{E|C} \; (\uy_E - {\hat\mu}_{E|C} ) / {n_E},
	\eeq
	\beq
	{\hat\mu}_{E|C} =F_E \ubhat_E \spp V^t_{CE}\; V^{-1}_{CC} \;({\uy}_C -  F_C \ubhat_C ).
	\eeq
\end{linenomath}

\subsection*{A.4 Test functions for toy model study}

\vspace{.2 cm} Test function 1: $nC=nE=30$ for all five test functions.
\begin{linenomath}
	\begin{eqnarray*} 
		&Y(\tau ,x) = \tau_{1} \exp( \tau_{2} +x_{1} )+ \tau_{1} x_{2}^{2} - \tau_{2} x_{3}^{2} \\
		&Computer~~ data:  T_{1} \sim U(0,5) ,\; T_{2} \sim U(0,4), \;  x_{1} \sim  U(-3, 3), \\  &x_{2} \sim U(-3, 3), \;  x_{3} \sim U(0, 6) \\
		&Experimental~~ data:   \tau_{1} =2 ,\;  \tau_{2} = 2, \;  \sigma_{E}^{2} =1.
\end{eqnarray*} \end{linenomath}

\vspace{.2 cm} Test function 2:
\begin{linenomath}
	\begin{eqnarray*} 
		&Y(\tau ,x)= \tau_{1}\exp{(\tau_{2}+ x_{1} + \tau_{3})}  + \tau_{1}\tau_{3}x_{2} ^{2} - \tau_{2}x_{3}^{2} - \tau_{3}\log{(x_{4})} \\
		&Computer ~~ data: T_{1} \sim U(0, 5) ,\;  T_{2} \sim U(0,4), \;  T_{3} \sim U(1,5), \\
		& x_{1} \sim U(-3, 4),\;  x_{2} \sim U(-3, 3),\;  x_{3} \sim U(0, 6), \;  x_{4} \sim U(1,5) \\
		& Experimental~~ data: \tau_{1} =2 ,\;  \tau_{2} = 1,\;  \tau_{3} = 3,\;  \sigma_{E}^{2} =1.
\end{eqnarray*} \end{linenomath}

\vspace{.2 cm} Test function 3:
\begin{linenomath}
	\begin{eqnarray*} 
		&Y(\tau ,x)= \tau_{1} \exp(|x_{1} +x_{2}|) + \tau_{2}(x_{3} +1.2x_{4} +1)/2.5
		+ \tau_{2} 3\cos( x_{2} + x_{3}) \\
		&Computer ~~data:  T_{1} \sim U(0, 4) , \; T_{2} \sim U(1,4),\;   x_{1} \sim U(-0.5, 1.5), \\
		&x_{2} \sim U(-0.5, 0.5),\;  x_{3} \sim U(-0.5, 1.5), \; x_{4} \sim U(-0.5, 0.5) \\
		&Experimental ~~data: \tau_{1} =2 ,\;  \tau_{2} = 3,\;  \sigma_{E}^{2} =0.1.
\end{eqnarray*} \end{linenomath}

\vspace{.2 cm} Test function 4:
\begin{linenomath}
	\begin{eqnarray*} 
		&Y(\tau,x)= \tau_{1} x_{1}(x_{2} -x_{3})  /  \log(\frac{x_{4}}{x_{5}}) \left( 1+ \frac{ \tau_{2}x_{1}x_{6}}{\log(x_{4}/x_{5})x_{2}^{2}x_{7}}+\frac{x_{1}}{x_{8}} \right) \\
		&Computer~~ data: T_{1} \sim U(5, 8) ,\;  T_{2} \sim U(1,3),\;  x_{1} \sim U(6370, 115600), \\
		& x_{2} \sim U(990, 1110), \;  x_{3} \sim U(700, 820),\;   x_{4} \sim U(100, 50000),  \;  x_{5} \sim U(0.05, 0.15),  \\
		&  x_{6} \sim U(1120, 1680),\;  x_{7} \sim U(9855, 12045),\;  x_{8} \sim U(63.1, 116) \\
		&Experimental~~ data: \tau_{1} =2\pi ,\;  \tau_{2} = 2,\;  \sigma_{E}^{2} =2.
\end{eqnarray*} \end{linenomath}

\vspace{.2 cm} Test function 5:
\begin{linenomath}
	\begin{eqnarray*} 
		&Y(\tau ,x)= \tau_{1}x_{1}^{2} + \tau_{2}x_{2} + \tau_{3}cos(x_{3}\pi) +\tau_{4}sin(x_{4}\pi)   \\
		&Computer ~~ data: T_{1} \sim U(0, 5) ,\;  T_{2} \sim U(0,5), \;  T_{3} \sim U(0,7), \;  T_{4} \sim U(0,5) \\
		& x_{1} \sim U(0, 3),\;  x_{2} \sim U(0, 3),\;  x_{3} \sim U(0, 2), \;  x_{4} \sim U(0,2) \\
		& Experimental~~ data: \tau_{1} =1 ,\;  \tau_{2} = 2,\;  \tau_{3} = 3,\;  \tau_{4} = 2,\;  \sigma_{E}^{2} =4. \\
\end{eqnarray*}  \end{linenomath}

The test function 4 was used in Morris and Mitchell (1995), which has a physical
interpretation that $y_C$ represents steady-state flow of water through a
borehole between two aquifers.

In each test function, $n_C$ values
of $(\uc, x)$ were selected as inputs for the ``computer code", that is, the
function $Y(\uc, x)$ is evaluated at $n_C$ values. The inputs were chosen to be
well spread around a reasonable space known to potentially contain the true
parameter value. $n_C$ computer data points and $n_E$
experimental data points were generated by using random Latin-hypercube
designs, except the $\csta$ values were used instead of $\uc$. In real
application situation, one would consider more sophisticated designs such as
data-adaptive sequential optimal experiments as reported in the next section.

\subsection*{A.5 Sequential designs in tuning a nuclear fusion simulator}

For given a Gaussian process model, the
A-optimal design is obtained by minimizing the integrated mean
squared error of prediction (MSEP) with respect to a design $S$ (Sacks et al. 1989),
\begin{linenomath}
	\beq \label{imsep} IMSE_S \;(\hat Y(x)) = \int_Q
	MSEP(\hat Y(x))\; d\mu(x) ,
	\eeq \end{linenomath}
where $Q$ is the design region,
and $\mu$ is a ``weight function" which may be the empirical
measure of uniformly distributed random points. Note that neither
$MSEP$ nor $IMSE$ depend on the unknown parameters $\beta$ and
$\sz2$,  but depend on $\theta,~\gc2$ and design
$S$. This makes it possible to design an experiment (for
specified values of $\theta$ and $\gc2$) before taking the data.

Because $\theta$ is generally not available for the initial design
stage, in our example, we used an rough estimate of $\theta$ based
on a previous similar work given by a Baldur specialist.
Initially we found 10 optimal design points for
eight variables ($\uc, ~P,~I,~N,B$) which minimize $IMSE$ over the
design region $Q$, with $\theta=.5$ and $\gc2=.001$.

The following is the data-adaptive sequential optimal design
procedure that we have used in this study.
\begin{description}
	\item[Step 1.] Collect computer observations  based on the given optimal
	design ($n_1$ points, say).
	\item[Step 2.] Find an appropriate model and estimates of parameters
	($\theta,~\beta,~\sz2,~\gc2$) from the computer data.
	\item[Step 3.] Check the MMSE, and stop constructing the next stage design if
	MMSE is smaller than a preassigned target value. Otherwise, go to
	the next step.
	\item[Step 4.] Use the estimates and model found in Step 2 to
	choose the next stage optimal design ($n_2$ points, say) under the
	condition that the previous design is given (i.e, update $n_2$
	more points to the previous design to make $n_1 + n_2$ points).
	\item[Step 5.] Collect ($n_2$) more observations, and go to Step 2.
\end{description}
The MMSE (maximum mean squared error of prediction) is defined as
\beq MMSE_S \;(\hat Y(x)) = {Max \atop {x_i \in Q}}~ MSEP \;( \hat
Y(x_i )),
\eeq
where $x_i,~ i=1,2,\cdots,K$, are d-dimensional
random vectors. Note that the MMSE is used here as a measure of
accuracy of a given prediction model.

\small
\renewcommand{\baselinestretch}{1.0}

\begin{table}[htbp]
	
	\caption{Result of test functions 1, 3 and 4 with Models 1 and 2.
		The standard deviation (SD) computed from 30 Latin-hypercube repetitions is given in parentheses.}  \vspace{.3 cm}
	\begin{tabular}{C{1.2cm}C{1.5cm}cc C{2.5cm}C{2.5cm}C{2.5cm}C{1.4cm}}
		\hline
		Test function& True values & model& method & Average of $\hat{\tau_{1}}$ (SD)  & Average of $\hat{\tau_{2}}$ (SD)
		& Average distance to the true value (SD) & MSE \\
		\hline
		1 & $\tau_1 =2,~~ \tau_2 = 2$ & Model 1 & ANLSE  & 1.609 (0.207) & 1.773 (0.246) & 0.527 (0.168) & 0.381\\
		&& & SMLE  & 1.790 (0.245) & 1.758 (0.275) & 0.462 (0.147) & 0.349\\
		&& & Max-min & 2.296 (0.135) & 1.966 (0.224) & 0.377 (0.109) & 0.211\\
		\hline
		1 & $\tau_1 =2,~~ \tau_2 = 2$  & Model 2& ANLSE & 1.448 (0.249) & 1.602 (0.351) & 0.749 (0.288) &  0.746 \\
		&& &SMLE & 2.228 (0.409) & 1.465 (0.267) & 0.703 (0.280) &  0.732 \\
		&& &Max-min  & 2.382 (0.399) & 2.151 (0.575) & 0.718 (0.342) & 1.005 \\
		\hline
		
		\hline
		3 & $\tau_1 =2,~~ \tau_2 = 3$ & Model 1& ANLSE  & 2.000(0.276)  & 2.984(0.264)  & 0.370(0.089) & 0.283 \\
		&& &SMLE  &2.038(0.217)  & 2.980(0.183) & 0.249(0.136) & 0.143  \\
		&& & Max-min  & 2.009(0.228)  & 2.985(0.194)  & 0.258(0.146) & 0.146 \\
		\hline
		4 & $\tau_1 =2\pi ,~~ \tau_2 = 2$ & Model 1& ANLSE  & 5.990 (0.442) & 2.078 (0.140) & 0.494 (0.239) & 0.460 \\
		&& &SMLE  & 6.016 (0.459) & 2.055 (0.135) & 0.485 (0.251) & 0.464 \\
		&& &Max-min & 6.120 (0.426) & 2.109 (0.122) & 0.420 (0.240) & 0.373 \\
		\hline
	\end{tabular}%
	\label{tab_test1}%
\end{table}%

\begin{table}[htbp]
	\centering
	\caption{Result of test function 2 with Model 1 when $\tau_{1}=2$, $\tau_{2}=1$, $\tau_{3}=3$. The standard deviation (SD) computed from 30 Latin-hypercube repetitions is given in parentheses.}
	\vspace{.3 cm}
	\begin{tabular}{cC{2.5cm}C{2.5cm}C{2.5cm}C{2.5cm} C{2cm}}
		\hline
		method & Average of $\hat{\tau_{1}}  (SD) $ & Average of $\hat{\tau_{2}}  (SD)$ & Average of $\hat{\tau_{3}}  (SD)$  &  Average distance to the true value  (SD)& MSE  \\
		\hline
		ANLSE & 1.801 (0.324) & 1.349 (0.225) & 3.001 (0.319) & 0.598 (0.232)& 0.615 \\
		SMLE  & 1.737 (0.565) & 1.080 (0.527) & 2.957 (0.519) & 0.873 (0.398) & 1.637\\
		Max-min  & 1.706 (0.293) & 0.842 (0.127) & 3.003 (0.435) & 0.535 (0.327) & 0.577\\
		\hline
	\end{tabular}%
	\label{tab_test2}%
\end{table}%

\begin{table}[htbp]
	\centering
	\caption{Result of test function 5 with Model 1 when $\tau_{1}=1$, $\tau_{2}=2$, $\tau_{3}=3$, $\tau_{4}=2$. The standard deviation (SD) computed from 30 repetitions is given in parentheses.}
	\vspace{.3 cm}
	\begin{tabular}{cC{2cm}C{2cm}C{2cm}C{2cm}C{2cm}C{2cm}}
		\hline
		method & Average of $\hat{\tau_{1}}  (SD) $ & Average of $\hat{\tau_{2}}  (SD)$ & Average of $\hat{\tau_{3}}  (SD)$ & Average of $\hat{\tau_{4}}  (SD)$  &  Average distance to the true value  (SD)& MSE  \\
		\hline
		ANLSE & 0.935 (0.237) & 1.753 (0.396) & 3.044 (0.908) & 1.951 (0.610)& 1.025 (0.628)&  2.460 \\
		SMLE & 0.845 (0.291) & 1.975 (0.758) & 2.875 (0.532) & 1.852 (0.607)& 1.069 (0.437)&  2.453 \\
		Max-min  & 0.562 (0.261) & 2.155 (0.472) & 3.087 (0.549) & 2.049 (0.560)& 0.955 (0.291)&1.818 \\
		\hline
	\end{tabular}%
	\label{tab_test5}%
\end{table}%

\begin{table}[htbp]
	\centering
	\caption{Result of test function 2 with Model 2 when $\tau_{1}=2$, $\tau_{2}=1$, $\tau_{3}=3$. The standard deviation (SD) computed from 30 repetitions is given in parentheses.}
	\vspace{.3 cm}
	\begin{tabular}{cC{2cm}C{2cm}C{2cm}C{2cm}C{2cm}}
		\hline
		method & Average of $\hat{\tau_{1}}  (SD) $ & Average of $\hat{\tau_{2}}  (SD)$ & Average of $\hat{\tau_{3}}  (SD)$  &  Average distance to the true value  (SD)& MSE  \\
		\hline
		ANLSE & 1.715 (0.244) & 1.107 (0.322) & 3.034 (0.344) & 0.570 (0.209)&  0.606 \\
		SMLE & 1.912 (0.347) & 0.901 (0.381) & 2.858 (0.416) & 0.585 (0.353)&  0.781 \\
		Max-min  & 1.768 (0.231) & 1.106 (0.314) & 3.054 (0.283) & 0.511 (0.176)& 0.493 \\
		\hline
	\end{tabular}%
	\label{tab_test4_m2}%
\end{table}%

\begin{table}[htbp]
	\centering
	\caption{Result of test function 5 with Model 2 when $\tau_{1}=1$, $\tau_{2}=2$, $\tau_{3}=3$, $\tau_{4}=2$. The standard deviation (SD) computed from 30 repetitions is given in parentheses.}
	\vspace{.3 cm}
	\begin{tabular}{cC{2cm}C{2cm}C{2cm}C{2cm}C{2cm}C{2cm}}
		\hline
		method & Average of $\hat{\tau_{1}}  (SD) $ & Average of $\hat{\tau_{2}}  (SD)$ & Average of $\hat{\tau_{3}}  (SD)$ & Average of $\hat{\tau_{4}}  (SD)$  &  Average distance to the true value  (SD)& MSE  \\
		\hline
		ANLSE & 0.874 (0.250) & 1.852 (0.414) & 2.765 (0.846) & 1.729 (0.672)& 1.149 (0.456)&  2.721 \\
		SMLE & 1.008 (0.515) & 1.895 (0.895) & 2.740 (0.475) & 1.816 (0.953)& 1.398 (0.530)&  4.154 \\
		Max-min  & 0.600 (0.230) & 2.319 (0.549) & 3.114 (0.749) & 2.104 (0.407)& 1.095 (0.362)& 2.280 \\
		\hline
	\end{tabular}%
	\label{tab_test5_m2}%
\end{table}%

\begin{table}[htbp]
	\centering
	\caption{Maximum likelihood estimates of the parameters of a Gaussian process Model 1 obtained using both computer and experimental data for PDX nuclear fusion.}
	\vspace{.3 cm}
	\begin{tabular}{llc}
		\hline
		Symbol & Description & Estimates \\
		\hline
		$n_E$ & Sample size of experiment data  & 42 \\
		$n_C$  & Sample size of computer data  & 64 \\
		$\theta$ & Parameter for covariance & 0.980  \\
		$\beta_0$ & Regression coefficient (intercept) & 0.025  \\
		$\beta_1$ & Regression coefficient for $T_1$ & -0.027  \\
		$\beta_2$ & Regression coefficient for $T_2$ & -0.010  \\
		$\beta_3$ & Regression coefficient for $T_3$ & 0.001  \\
		$\beta_4$ & Regression coefficient for $T_4$ & -0.015  \\
		$\beta_5$ & Regression coefficient for \it{P} & -0.031  \\
		$\beta_6$ & Regression coefficient for \it{I} & 0.009  \\
		$\beta_7$ & Regression coefficient for \it{N} & -0.012  \\
		$\beta_8$ & Regression coefficient for \it{B} & 0.017  \\
		$\sigma^2$ & Variance of \it{Y} & 1.506E-04  \\
		$\gamma_E$ & Variance of $\see2/\sigma^2$ & 0.454  \\
		\hline
	\end{tabular}%
	\label{tab_tok_mle}%
\end{table}%

\begin{figure}[tbh]
	\centerline{\includegraphics[width=9cm,height=6cm]{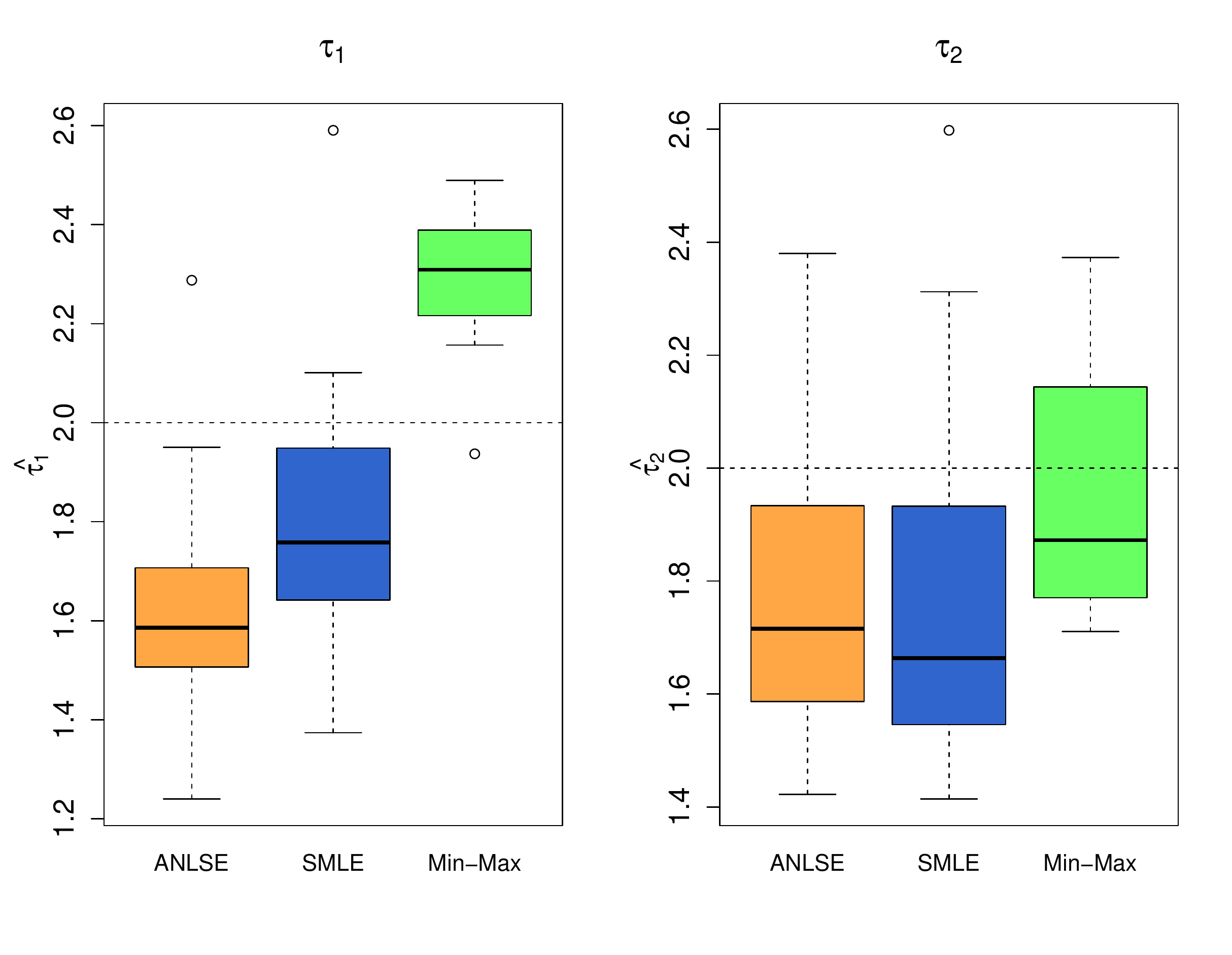}}
	\caption{Box plot of the tuning parameter estimates ($\hat{\tau}$) in the test function 1, obtained from 30 Latin-hypercube design experiments
		using a Gaussian process Model 1.  The horizontal dotted line denotes the true value.
		\label{test1_box}}\vspace{.2cm}
\end{figure}

\begin{figure}[tbh]
	\centerline{\includegraphics[width=10cm,height=6cm]{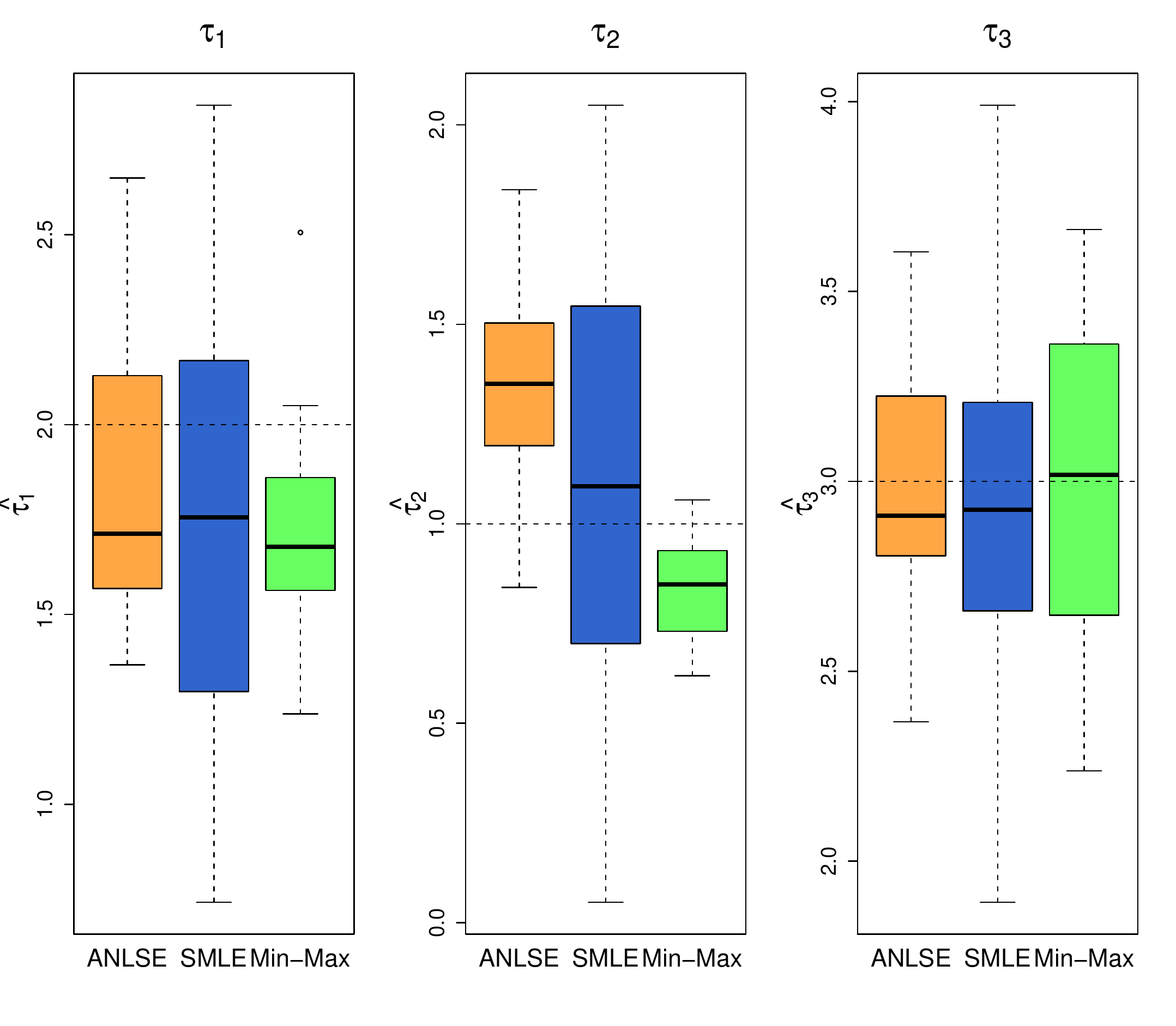}}
	\caption{Same as Figure S1 but the test function 2.
		\label{test2_box}}\vspace{.2cm}
\end{figure}

\begin{figure}[tbh]
	\centerline{\includegraphics[width=9cm,height=6cm]{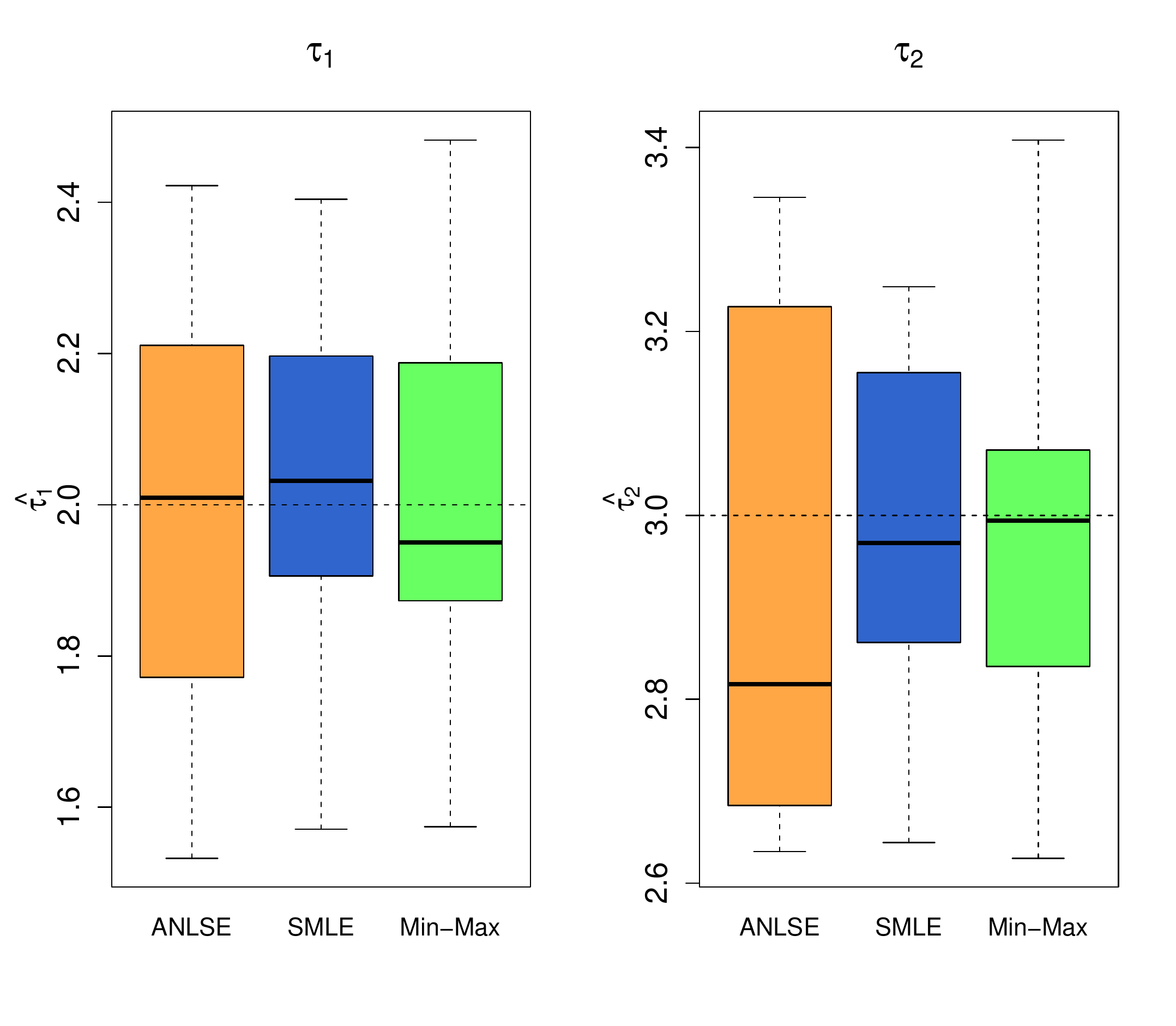}}
	\caption{Same as Figure S1 but the test function 3.
		\label{test3_box}}\vspace{.2cm}
\end{figure}

\begin{figure}[tbh]
	\centerline{\includegraphics[width=9cm,height=6cm]{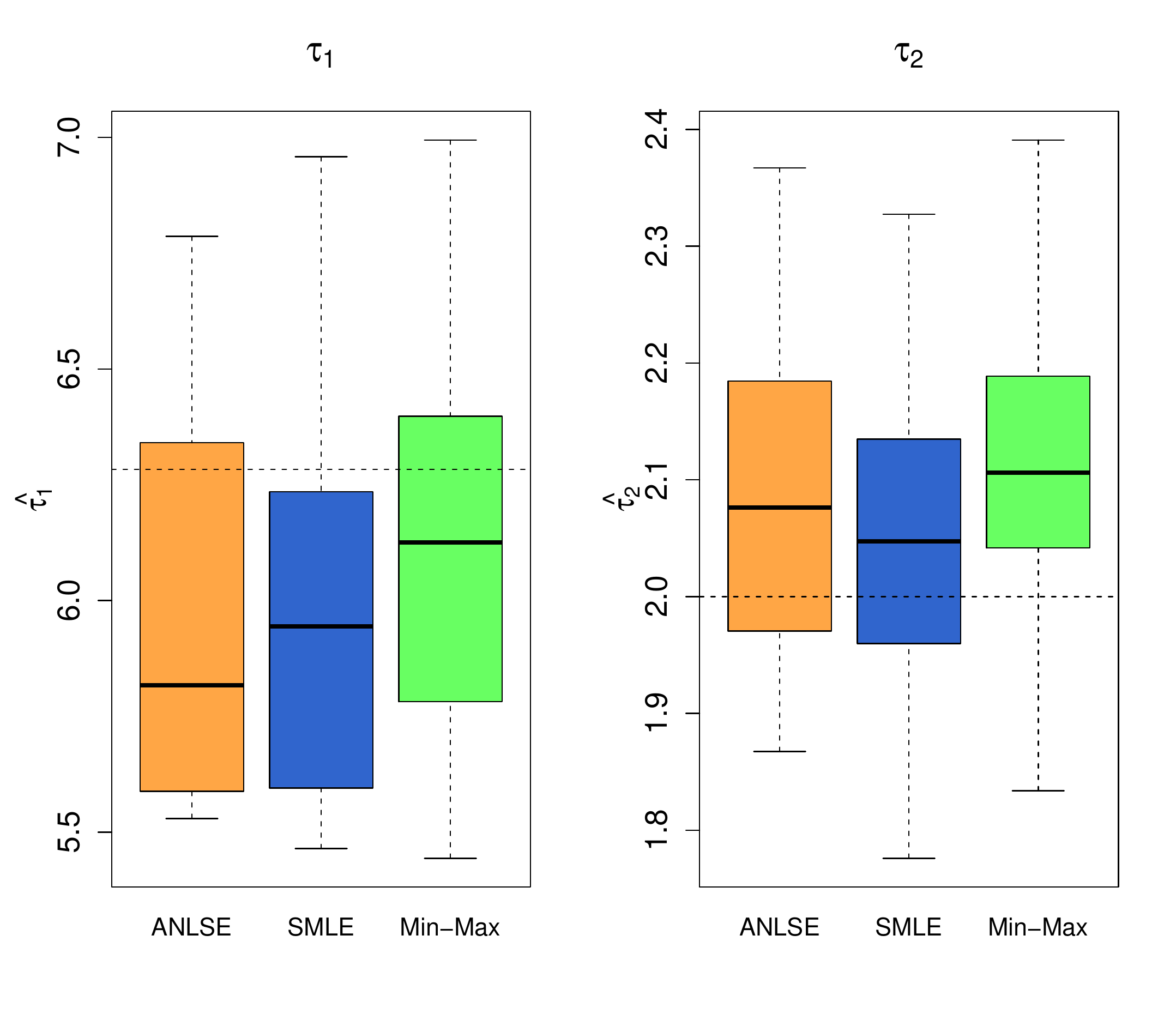}}
	\caption{Same as Figure S1 but the test function 4.
		\label{test4_box}}\vspace{.2cm}
\end{figure}

\begin{figure}[tbh]
	\centerline{\includegraphics[width=11cm,height=7cm]{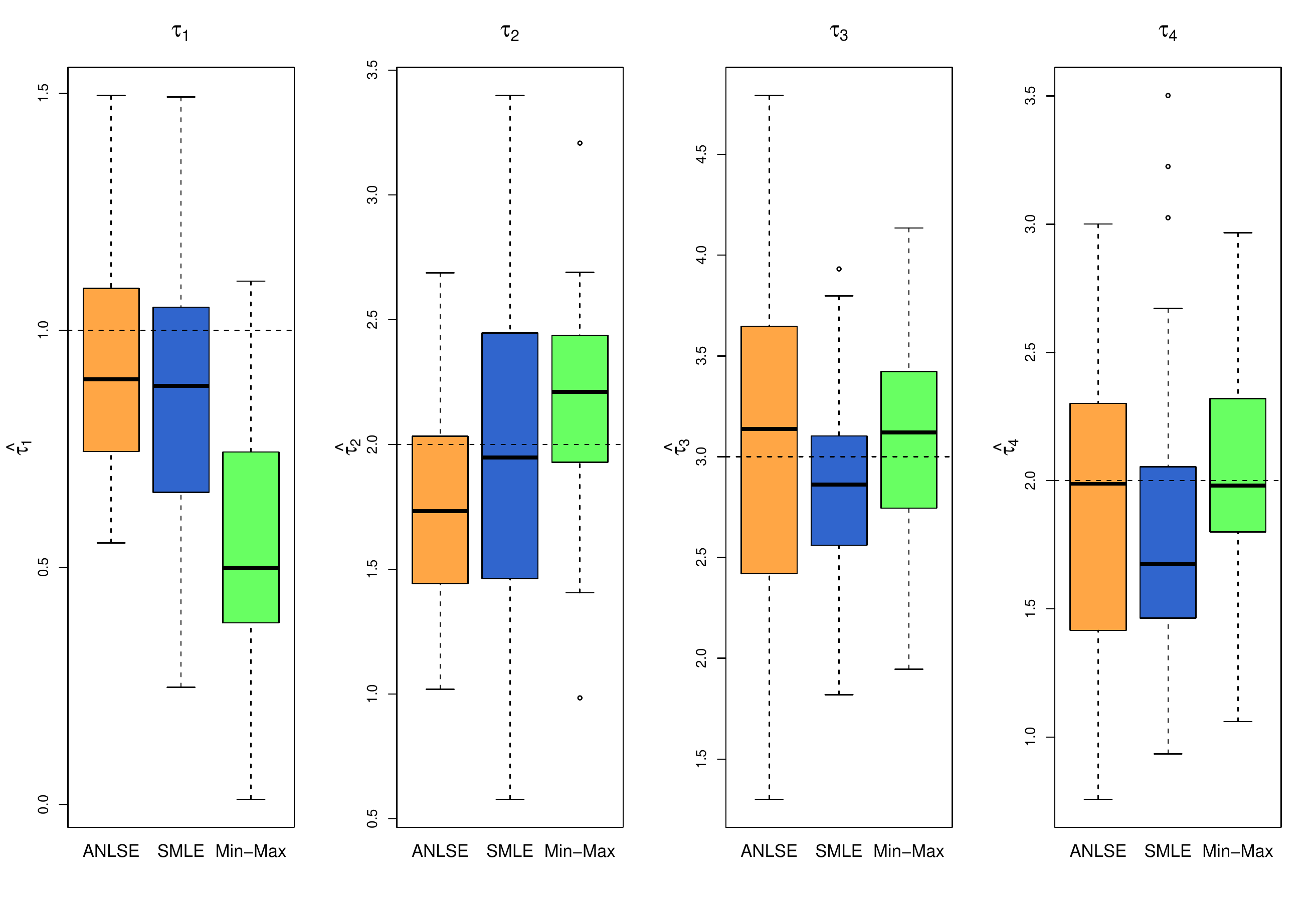}}
	\caption{Same as Figure S1 but the test function 5.
		\label{test5_box}}\vspace{.2cm}
\end{figure}

\begin{figure}[tbh]
	\centerline{\includegraphics[width=11cm,height=8cm]{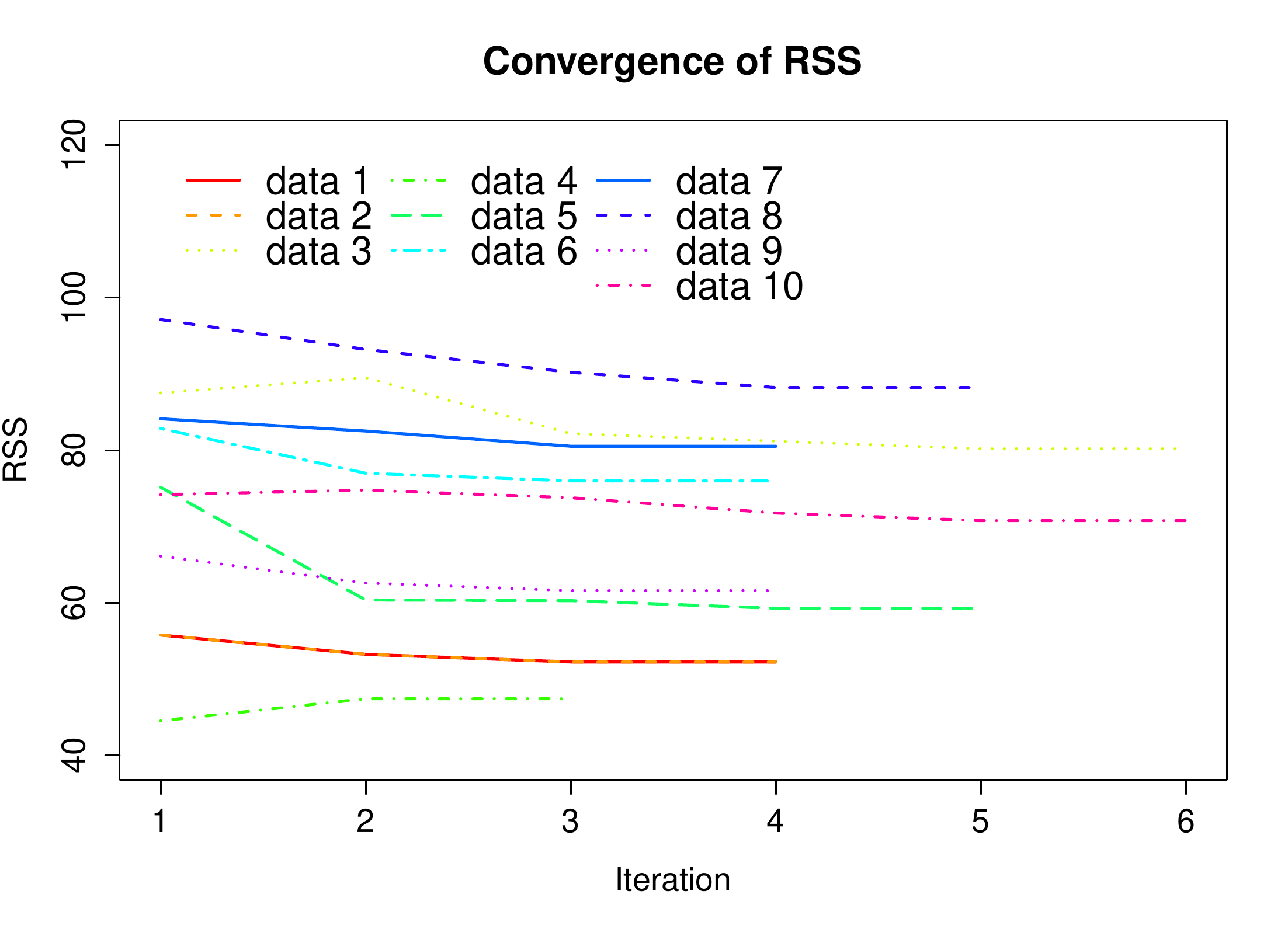}}
	\caption{Convergence of the Max-min algorithm as iteration increases for 10 Latin-hypercube design experiments in the test function 1.
		\label{converge}}\vspace{.2cm}
\end{figure}

\end{document}